\documentclass{article} 
\usepackage{preprintbox}

\usepackage{amsmath,amsfonts,bm}

\def\eqref#1{equation~\ref{#1}}

\def\1{\bm{1}}

\DeclareMathAlphabet{\mathsfit}{\encodingdefault}{\sfdefault}{m}{sl}
\SetMathAlphabet{\mathsfit}{bold}{\encodingdefault}{\sfdefault}{bx}{n}

\usepackage{hyperref}
\usepackage{url}

\usepackage{graphicx}
\usepackage{blindtext}
\usepackage{booktabs}

\usepackage{tabularx}
\usepackage{array}
\usepackage{enumitem}
\usepackage{graphicx}
\usepackage{subcaption}
\usepackage[dvipsnames]{xcolor}
\usepackage{ragged2e}
\usepackage{fontawesome5}
\usepackage{multirow}
\usepackage{array}
\usepackage{graphicx}
\usepackage{booktabs}
\newcommand{\mpmag}{\ensuremath{\mathrm{MP}}}
\newcommand{\chgmag}{\ensuremath{\mathrm{CG}}}
\newcommand{\pos}[1]{\textcolor{green!50!black}{$+#1$}}
\newcommand{\nega}[1]{\textcolor{red!70!black}{$-#1$}}
\newcolumntype{C}[1]{>{\centering\arraybackslash}m{#1}}

\newcommand{\scmark}{{\tiny\cmark}}
\newcommand{\crit}[1]{%
    {\color{red!70!black}\small\bfseries C#1}%
}
\newcounter{backmatter}
\newcommand{\backmattersection}[1]{%
  \stepcounter{backmatter}%
  \section*{\large #1}%
  \pdfbookmark[1]{#1}{backmatter.\thebackmatter}%
}
\newcommand{\lcrit}[1]{%
    {\color{red!70!black}\normalsize\bfseries C#1}%
}

\DeclareRobustCommand{\critref}[1]{%
    \hyperref[tab:prior_scope]{%
        \textcolor{red!70!black}{\textbf{C#1}}%
    }%
}

\newcommand{\mlmark}{%
  \makebox[0pt][c]{%
    \raisebox{2.1ex}{\tiny \textcolor{blue}{\textbf{mm}}}%
  }%
  \scmark%
}
\newcommand{\sxmark}{{\scriptsize$\times$}}

\usepackage[toc,page,header]{appendix}
\usepackage{titletoc}

\title{Complete Neural Electronic Initialization Accelerates Materials DFT}

\IfFileExists{latexml.sty}{\usepackage{latexml}}{\newif\iflatexml}
\newcommand{\eqmark}{\textsuperscript{*}}

\iflatexml
  \author{{Felix Ærtebjerg\eqmark, Jonas Elsborg\eqmark, Arghya Bhowmik}
          \\ {Department of Energy Conversion and Storage, Technical University of Denmark}
          \\ {\eqmark Equal contribution.}}
\else
  \author{Felix Ærtebjerg$^{*}$, Jonas Elsborg$^{*}$, Arghya Bhowmik}
  \affiliation{Department of Energy Conversion and Storage, Technical University of Denmark}
  \contribution[*]{Equal contribution.}
\fi

\abstract{
   We present the first complete machine learning method for accelerating plane-wave density functional theory (DFT) in materials under the projector augmented wave (PAW) formalism. We formalize seven criteria that a \textit{Complete Neural Electronic Initializer} must satisfy for practical end-to-end PAW DFT acceleration. Applying these to prior work reveals two structure-dependent components, augmentation occupancies and spin initialization, whose absence prevents existing acceleration methods from providing complete reference-free initialization. We show that omitting these components can eliminate or reverse the acceleration obtained via models that only predict the smooth valence density. We satisfy the missing requirements by introducing AugNet, a general equivariant model for PAW augmentation occupancies, and the first general spin density model for materials, which predicts the smooth spin-difference density and spin-difference PAW augmentation occupancies using predicted magnetic moments to constrain the global magnetic state. Combined with existing valence density models, our full method satisfies all seven criteria and forms a fully reference-free electronic initializer for materials DFT, requiring no electronic quantities from a converged target calculation. We show that perfect initialization could cut PAW DFT wall time by $40$-$52\%$, and our method recovers up to $62\%$ of this saving, reducing end-to-end DFT wall time by up to $\sim25\%$ on unseen structures while preserving converged energies.}

\correspondence{FÆ: \email{felar@dtu.dk}, AB: \email{arbh@dtu.dk}}
\code{https://github.com/aerte/neural_paw_dft}

\usepackage{pifont}

\definecolor{progreen}{HTML}{2E7D32}
\definecolor{conred}{HTML}{C62828}

\renewcommand{\arraystretch}{1.2}

\begin{document}

\maketitle

\begin{figure}[h!]
    \centering
    \includegraphics[width=1.0\linewidth]{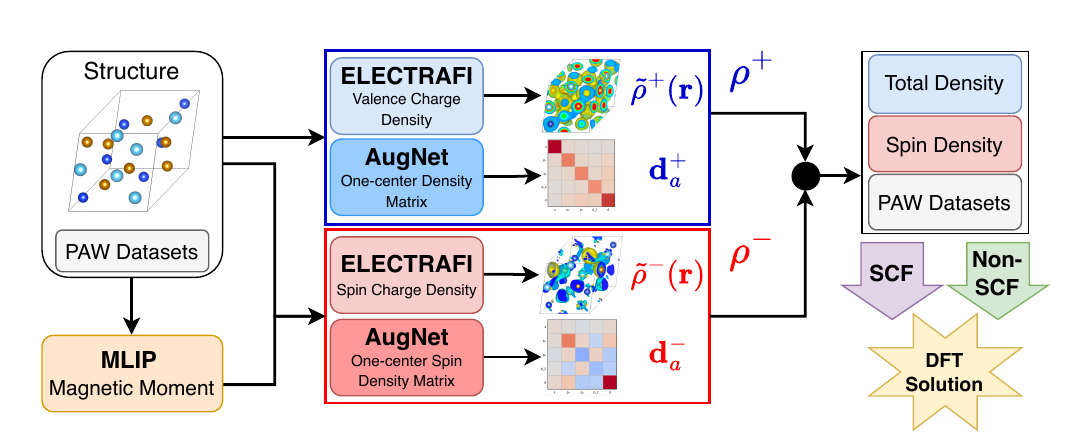}
    \caption{
    Workflow for end-to-end PAW DFT acceleration, illustrated using the models
    employed in this work. For a given atomic structure and PAW setup, the fixed
    PAW datasets define the frozen core and basis information, while a Complete
    Neural Electronic Initializer provides the three structure-dependent components
    defined in Table~\ref{tab:prior_scope}: the smooth valence density
    $\tilde{\rho}^{+}$, PAW augmentation occupancies $\mathbf d_a^{+}$, and
    spin/magnetic initialization. ELECTRAFI predicts $\tilde{\rho}^{+}$, AugNet
    predicts $\mathbf d_a^{+}$, and the corresponding spin-dependent components
    $\tilde{\rho}^{-}$ and $\mathbf d_a^{-}$ are predicted by spin-ELECTRAFI and
    spin-AugNet, with magnetic moment predictions constraining the global magnetic
    state. Together, these components enable complete reference-free initialization
    of PAW DFT.
    }
    \label{fig:init_workflow}
\end{figure}

\section{Introduction \& Motivation}
Density functional theory (DFT) is a central computational tool in materials science, chemistry, and condensed matter physics that enables modeling of atomic and electronic properties across large chemical spaces~\citep{hohenberg1964inhomogeneous,kohn1965self,jain2013commentary,gavini2023roadmap}. DFT calculations account for up to 45\% of core hours on the UK ARCHER2 Tier-1 supercomputer and over 70\% of allocation time within the materials science sector at NERSC~\citep{riebesell2025framework}, and generating the 118 million inorganic structures in OMat24 required more than 400 million CPU core hours of DFT calculations~\citep{barros2026open}. At this scale, even single-digit percentage reductions in DFT cost translate to tens of millions of CPU core hours saved. The importance of DFT has only increased as large DFT datasets have become the basis for machine-learned interatomic potentials (MLIPs), universal atomistic models, and materials discovery pipelines~\citep{batzner20223,batatia2022mace,chen2022universal,deng2023chgnet,batatia2025foundation,qu2024importance,neumann2024orb}. 
For high-throughput computational materials science, plane-wave DFT with the projector augmented wave (PAW) formalism is the de facto standard. VASP is the dominant PAW implementation~\citep{blochl1994projector,kresse1999ultrasoft} and underlies many of the field's canonical materials datasets, including the Materials Project, OQMD, AFLOW, JARVIS-DFT, GNoME, and OMat24~\citep{jain2013commentary,kirklin2015open,curtarolo2012aflowlib,choudhary2020joint,merchant2023scaling,barros2026open}. In PAW DFT, the quality of the initial electronic quantities matters because better initialization can reduce the number of self-consistent field (SCF) cycles required for convergence. Charge density is one such input, and has motivated machine learning (ML) models that predict charge densities directly from atomic structure~\citep{jorgensen2022equivariant,kim2024gaussian,cheng2024equivariant,koker2024higher,fu2024recipe,elsborg2025electra,klockowfunction,elsborg2026electrafi}. Unlike MLIPs, which accelerate atomistic simulation by approximating the DFT potential energy surface, these models target the cost of obtaining the DFT solution itself, without replacing the underlying electronic structure model. However, most of these studies evaluate only density prediction accuracy, assumed to be a proxy for acceleration potential~\citep{kim2024gaussian,cheng2024equivariant,fu2024recipe,klockowfunction}. More importantly, even studies that evaluate DFT acceleration directly do not take into account that charge density is not the only required electronic input quantity and therefore does not constitute the complete PAW electronic state. The experiments are only feasible because crucial components are retained from converged reference calculations~\citep{jorgensen2022equivariant,koker2024higher,elsborg2025electra,elsborg2026electrafi}. Thus, such experiments isolate the quality of the learned smooth valence density, but do not represent a real ML-accelerated DFT workflow, since the retained converged quantities are unavailable for new calculations. A notable exception is~\citet{sunshine2023chemical}, who found no acceleration over VASP's default initialization using a predicted valence density and PAW augmentation occupancies from a zero-step VASP calculation, identifying augmentation and wavefunction initialization as remaining bottlenecks. Our experiments support this conclusion and show that even a converged valence density provides little acceleration when the remaining PAW components are poorly initialized.
\newcommand{\cmark}{\scalebox{2.0}{$\checkmark$}}
\newcommand{\xmark}{\scalebox{2.0}{$\times$}}
\renewcommand{\crit}[1]{%
    {\color{red!70!black}\small\bfseries C#1}%
}
\begin{table}[t!]
\centering
\caption{
Criteria for complete neural electronic initialization for accelerating materials DFT. \textbf{\textcolor{red!70!black}{C1-C3}} are structure-dependent quantities.
\textbf{\textcolor{red!70!black}{C4-C7}} refer to the required evaluation for demonstrating complete initialization. $\checkmark$ denotes a capability in the cited work, $\times$ denotes that it was not demonstrated. $\dagger$ denotes system-specific models: CJM~\citep{focassio2024covariant} is specific to MoS$_2$, while \citet{de2023nanosecond} and EAC-Net~\citep{qin2026eac} predict spin densities for Na$_3$V$_2$(PO$_4$)$_3$ and Fe, respectively, without demonstrating SCF acceleration. 
}
\label{tab:prior_scope}
\scriptsize
\setlength{\tabcolsep}{2.8pt}
\renewcommand{\arraystretch}{1.15}

\resizebox{\linewidth}{!}{
\begin{tabular}{lccccccc}
\toprule
&
\multicolumn{3}{c}{\textbf{Reference-free initialization components}}
&
\multicolumn{4}{c}{\textbf{Demonstrated evaluation}}
\\
\cmidrule(lr){2-4}
\cmidrule(lr){5-8}

\textbf{Method}
&
\shortstack{\textbf{Valence}\\\textbf{density}\\\lcrit{1}}
&
\shortstack{\textbf{PAW}\\\textbf{augmentation}\\\lcrit{2}}
&
\shortstack{\textbf{Spin / magnetic}\\\textbf{initialization}\\\lcrit{3}}
&
\shortstack{\textbf{Density}\\\textbf{accuracy}\\\lcrit{4}}
&
\shortstack{\textbf{SCF}\\\textbf{acceleration}\\\lcrit{5}}
&
\shortstack{\textbf{Component}\\\textbf{ablations}\\\lcrit{6}}
&
\shortstack{\textbf{Reference-free}\\\textbf{acceleration}\\\lcrit{7}}
\\
\midrule

GPWNO \citep{kim2024gaussian}
& \cmark & \xmark & \xmark
& \cmark & \xmark & \xmark & \xmark \\

InfGCN \citep{cheng2024equivariant}
& \cmark & \xmark & \xmark
& \cmark & \xmark & \xmark & \xmark \\

SCDP \citep{fu2024recipe}
& \cmark & \xmark & \xmark
& \cmark & \xmark & \xmark & \xmark \\

BOA \citep{klockowfunction}
& \cmark & \xmark & \xmark
& \cmark & \xmark & \xmark & \xmark \\

EdenGNN \citep{li2025efficient}
& \cmark & \cmark & \xmark
& \cmark & \xmark & \xmark & \xmark \\

\midrule

DeepDFT \citep{jorgensen2022equivariant}
& \cmark & \xmark & \xmark
& \cmark & \cmark & \xmark & \xmark \\

ChargE3Net \citep{koker2024higher}
& \cmark & \xmark & \xmark
& \cmark & \cmark & \xmark & \xmark \\

ELECTRA \citep{elsborg2025electra}
& \cmark & \xmark & \xmark
& \cmark & \cmark & \xmark & \xmark \\

ELECTRAFI \citep{elsborg2026electrafi}
& \cmark & \xmark & \xmark
& \cmark & \cmark & \xmark & \xmark \\

\midrule

NASICON model \citep{de2023nanosecond}
& \cmark$^{\dagger}$
& \xmark
& \cmark$^{\dagger}$
& \cmark$^{\dagger}$
& \xmark
& \xmark
& \xmark \\

CJM \citep{focassio2024covariant}
& \cmark$^{\dagger}$
& \cmark$^{\dagger}$
& \xmark
& \cmark$^{\dagger}$
& \xmark
& \xmark
& \xmark \\

EAC-Net \citep{qin2026eac}
& \cmark
& \xmark
& \cmark$^{\dagger}$
& \cmark
& \xmark
& \xmark
& \xmark \\

\midrule

\textbf{\large This work}
& \cmark
& \cmark
& \cmark
& \cmark
& \cmark
& \cmark
& \cmark
\\

\bottomrule
\end{tabular}
}
\end{table}
\\

To clarify the distinction, we formalize seven criteria that must be met to demonstrate practical, reference-free ML-driven acceleration of PAW DFT in materials. We refer to such a method as a \textit{Complete Neural Electronic Initializer}. Such a method must provide three structure-dependent initialization components:
\textbf{\critref{1}}, the smooth valence density $\tilde{\rho}^{+}$;
\textbf{\critref{2}}, the PAW augmentation occupancies $\mathbf d_a^{+}$;
and \textbf{\critref{3}}, spin/magnetic initialization, including the
spin-dependent components $\tilde{\rho}^{-}$ and $\mathbf d_a^{-}$. A complete demonstration must also establish \textbf{\critref{4}}, density accuracy; \textbf{\critref{5}}, SCF acceleration; \textbf{\critref{6}}, controlled component ablations; and \textbf{\critref{7}}, reference-free end-to-end wall time reduction including ML inference. Prior methods satisfy two or at most three of these seven criteria, and no prior method has been published that jointly addresses general PAW augmentation, magnetic initialization, component importance, and corresponding reference-free end-to-end acceleration. We summarize our criteria jointly with the state of the field in Table~\ref{tab:prior_scope}.

\paragraph{Contributions.}
We present the first Complete Neural Electronic Initializer, satisfying all criteria in Table~\ref {tab:prior_scope}. Specifically, we address the three criteria no prior general method satisfies, magnetic initialization (\critref{3}), controlled component ablations (\critref{6}), and reference-free end-to-end acceleration (\critref{7}), and evaluate predicted PAW augmentation (\critref{2}) as an SCF initializer. Our contributions are:
\begin{enumerate}
    \item \textbf{We establish the requirements for reference-free initialization.}
    We show that valence, augmentation, and spin initialization components must all be treated explicitly for practical SCF acceleration, and identify the contribution of each.

    \item \textbf{We introduce AugNet, a general equivariant model for PAW
    augmentation occupancies.}
    AugNet satisfies \critref{2} by predicting structured one-center augmentation coefficients across diverse materials, elements, and PAW schemas, including the spin-difference channel, and is to our knowledge the first such model evaluated as an SCF initializer.

    \item \textbf{We introduce the first general model for spin density prediction.}
    Using a charge-informed CHGNet model to constrain the ELECTRAFI model's density prediction, we enable direct prediction of spin difference densities, satisfying \critref{3}.

    \item \textbf{We demonstrate the first reference-free ML acceleration
    of PAW DFT.}
    By combining all components, we present a method that satisfies all requirements in Table~\ref{tab:prior_scope} and reduces total DFT wall time by up to $\sim25\%$ on unseen structures.
\end{enumerate} 

\section{Background \& Related Work}\label{section:background}
\paragraph{DFT \& PAW.}
Density functional theory (DFT) is the standard first-principles framework for electronic structure calculations in materials ~\citep{hohenberg1964inhomogeneous,kohn1965self}. In periodic systems, Kohn-Sham DFT is commonly solved in a plane-wave basis through self-consistent field (SCF) iteration of the electronic density~\citep{payne1992iterative,kresse1996efficient}. The projector augmented wave (PAW) method~\citep{blochl1994projector} enables efficient plane-wave DFT by replacing the rapidly varying all-electron wavefunctions near the nuclei with smooth pseudo wavefunctions, while restoring
the missing atom-centered information through one-center corrections. The PAW decomposition can be written as
\begin{equation}
    \rho^{+}(\mathbf r)
    =
    \tilde{\rho}^{+}(\mathbf r)
    +
    \sum_a
    \left[
        \rho^{a,+}(\mathbf r)
        -
        \tilde{\rho}^{a,+}(\mathbf r)
    \right],
    \label{eq:paw-density-decomp}
\end{equation}
where $\tilde{\rho}^{+}(\mathbf r)$ is the spin-summed smooth valence density
represented on the plane-wave grid, while $\rho^{a,+}-\tilde{\rho}^{a,+}$ restores the atom-centered all-electron information removed by the smoothing procedure. While the basis functions defining
$\rho^{a,+}-\tilde{\rho}^{a,+}$ are fixed by the PAW dataset, their
coefficients depend on the electronic state of the material. In
VASP~\citep{kresse1996efficient,kresse1999ultrasoft}, these structure-dependent
coefficients are stored as augmentation occupancies. 
Thus, a complete ML initialization method must predict the PAW augmentation occupancies to satisfy \critref{2}. EdenGNN~\citep{li2025efficient} predicts PAW augmentation occupancies for general non-magnetic materials, but uses them to build a non-self-consistent Hamiltonian rather than to initialize SCF, and does not treat spin. CJM also directly predicts PAW
augmentation occupancies, but is system-specific to MoS$_2$ and does not
evaluate SCF acceleration~\citep{focassio2024covariant}. Spin-polarized PAW
initialization additionally requires the smooth spin-difference density
$\tilde{\rho}^{-}$ and corresponding augmentation occupancies
$\mathbf d_a^{-}$. Prior materials spin density models are likewise
system-specific~\citep{de2023nanosecond,qin2026eac}, with no demonstration of
SCF acceleration. Further details on VASP representation, initialization procedure, and influence of individual components are in Appendix~\ref{app:baseline_scf}. We provide details on augmentation occupancies in Appendix~\ref{app:mace-aug}, and on spin-polarized and magnetic calculations in Appendix~\ref{app:magnetic-init}.

\paragraph{Charge density prediction.}
Nearly all existing charge density models predict the $\tilde{\rho}^{+}(\mathbf r)$ valence density term in Equation~\ref{eq:paw-density-decomp}. Models differ mainly in how they represent the map from atomic structure $\mathcal X={(Z_i,\mathbf R_i)}_{i=1}^N$ to $\tilde{\rho}^{+}(\mathbf r)$. The state-of-the-art in the field is ELECTRAFI~\citep{elsborg2025electra} and ChargE3Net~\citep{koker2024higher}. ELECTRAFI extends ELECTRA's~\citep{elsborg2025electra} floating Gaussians to materials by analytically transforming predicted floating Gaussians into reciprocal-space coefficients and reconstructing the density through inverse FFT~\citep{elsborg2026electrafi}. This avoids dense real-space neural evaluation and results in low inference cost. ChargE3Net achieves higher grid accuracy, but requires neural evaluation across the real-space grid and explicit periodic treatment~\citep{koker2024higher}. Its high inference cost therefore decreases the resulting wall time benefit~\citep{elsborg2026electrafi}. A broader overview of architectures, including related initialization methods outside the general periodic materials setting, is provided in Appendix~\ref{app:density_models}.

\paragraph{Requirements for complete neural electronic initialization in materials.}
ML models that predict only the smooth valence density
$\tilde{\rho}^{+}(\mathbf r)$ of Equation~\ref{eq:paw-density-decomp} can, at
most, satisfy \critref{1}, \critref{4}, and \critref{5}. Existing SCF acceleration studies largely follow the evaluation protocol introduced by~\citet{jorgensen2022equivariant}, in which only $\tilde{\rho}^{+}(\mathbf r)$ is replaced by an ML prediction. Their initialization is therefore effectively
\begin{equation}
    \rho_{\mathrm{init}}^{+}(\mathbf r)
    =
    \underbrace{\tilde{\rho}_{\mathrm{ML}}^{+}(\mathbf r)}_{\critref{1}}
    +
    \underbrace{
    \sum_a
    \left[
        \rho_{\mathrm{test}}^{a,+}(\mathbf r)
        -
        \tilde{\rho}_{\mathrm{test}}^{a,+}(\mathbf r)
    \right]
    }_{\text{converged augmentation from the same test structure}},
    \label{eq:prior-reference-dependence}
\end{equation}
with converged spin-dependent quantities likewise retained for spin-polarized calculations. These quantities are unavailable for a genuinely new calculation, so such methods do not satisfy \critref{2} or \critref{3}. Moreover, without isolating the contribution of these retained quantities they do not satisfy \critref{6}. The absence of \critref{2}, \critref{3}, and \critref{6} therefore precludes a reference-free end-to-end demonstration satisfying \critref{7}. We discuss these limitations in more detail in  Appendix~\ref{app:limitations}.
\section{Methods} \label{sec:method}
\paragraph{Valence and augmentation density.}
PAW augmentation contributions are strongly localized and atom-centered, whereas the smooth valence density is spatially extended and captures interatomic density~\citep{blochl1994projector,kresse1999ultrasoft}. We therefore model them separately. We use ELECTRAFI~\citep{elsborg2026electrafi} and ChargE3Net~\citep{koker2024higher} for the smooth valence density required by \critref{1}, and introduce AugNet below to model the PAW augmentation occupancy prediction required by \critref{2}.

\begin{figure}[t!]
    \centering
    \includegraphics[width=1.0\linewidth]{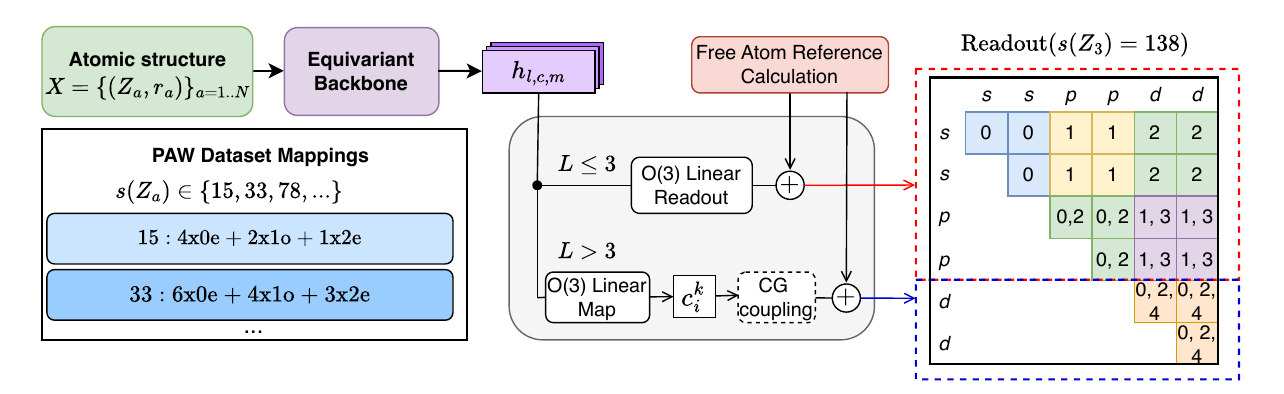}
    \caption{AugNet architecture. An equivariant backbone, schema-conditioned readout, and Clebsch-Gordan coupling predict on-site PAW coefficients. See Appendix~\ref{app:mace-aug} for details.}
    \label{fig:mace-aug-arch}
\end{figure}

\paragraph{AugNet: Augmentation occupancy prediction (\critref{2}).} PAW augmentation occupancies are the finite coefficient representation of the one-center correction $\rho^{a,+}-\tilde{\rho}^{a,+}$ in Equation~\ref{eq:paw-density-decomp}. They define a variable-schema equivariant prediction problem, see Appendix~\ref{app:wigner} for a proof. For atom $a$ with PAW schema $s_a=s(Z_a)$, the target space is
\begin{equation}
    \mathcal V_{s_a}
    =
    \bigoplus_L n_L^{(s_a)} D^L,
    \qquad
    F_\theta(\mathcal N_a,s_a):
    \mathcal N_a \mapsto \mathbf d_a^{+} \in \mathcal V_{s_a},
    \label{eq:augnet-schema}
\end{equation}
Here, $D^L$ is the $(2L+1)$-dimensional irrep of $SO(3)$ and
$\mathbf d_a^{+}=\{d^{LM,+}_{a,ij}\}_{ijLM}$ is the packed spin-summed
augmentation occupancy vector, where $i,j$ index PAW partial-wave channels and
$M=-L,\ldots,L$. Together with the fixed PAW basis,
\begin{equation}
    \rho^{a,+}(\mathbf r)-\tilde{\rho}^{a,+}(\mathbf r)
    =
    \sum_{ij,L,M}
    d^{LM,+}_{a,ij}\,
    B^{a,LM}_{ij}(\mathbf r),
    \label{eq:augnet-paw-reconstruction}
\end{equation}
where $B^{a,LM}_{ij}$ is fixed by the PAW dataset. Because both the
multiplicities $n_L^{(s_a)}$ and maximum $L$ depend on the PAW schema
($L\leq6$ in our data), AugNet must map atomic environments to
element-dependent output representations. We use a shared equivariant backbone and schema-conditioned readout. For the
spin-summed channel, AugNet predicts corrections to VASP's default augmentation occupancies based on  superposition of atomic densities (SAD),
\begin{equation}
    \widehat{\Delta\mathbf d}_a^{+}
    =
    \mathbf G_{s(Z_a)}
    \mathcal R(\mathbf h_a),
    \qquad
    \hat{\mathbf d}_a^{+}
    =
    \mathbf d_a^{+,\mathrm{SAD}}
    +
    \widehat{\Delta\mathbf d}_a^{+},
    \label{eq:augnet-map}
\end{equation}
where $\mathcal R$ is the shared equivariant readout applied to the atom-wise backbone representation $\mathbf h_a$, and $\mathbf G_{s(Z_a)}$ gathers the coefficients required by the PAW schema of element $Z_a$. Backbone-supported angular channels use equivariant linear maps, while higher-order PAW components are constructed by Clebsch-Gordan coupling,
\begin{equation}
    \widehat{\Delta d}^{LM,+}_{a,ij}
    =
    \sum_k w_{(ij,L),k}
    \sum_{m_i,m_j}
    C^{LM}_{\ell_i m_i,\ell_j m_j}
    c^{(k)}_{a,i,m_i}
    c^{(k)}_{a,j,m_j},
    \qquad
    L>L_{\mathrm{backbone}},
    \label{eq:augnet-cg}
\end{equation}
where $\ell_i,m_i$ and $\ell_j,m_j$ label the angular components of the
partial-wave channels, $C^{LM}_{\ell_i m_i,\ell_j m_j}$ are
Clebsch--Gordan coefficients, $k$ indexes learned projection channels, and
$c^{(k)}_{a,i,m_i}$ are learned equivariant projections. Figure~\ref{fig:mace-aug-arch} summarizes the architecture. 

AugNet is trained with a masked coefficient space loss over the valid PAW channels,
\begin{equation}
    \mathcal L_{\mathrm{Aug}}
    =
    \frac{
        \sum_{a,\alpha} m_{a,\alpha}
        \left(
            \hat d^{+}_{a,\alpha}-d^{+}_{a,\alpha}
        \right)^2
    }{
        \sum_{a,\alpha} m_{a,\alpha}
    },
    \label{eq:augnet-loss}
\end{equation}
where $\alpha$ indexes the $(i,j,L,M)$ coefficients. Implementation details are in Appendix~\ref{app:mace-aug}.

\paragraph{Spin \& magnetic modeling (\critref{3}).}
\critref{3} requires initializing the spin-dependent electronic state without access to a converged magnetization density. One option is to predict magnetic moments for each atom and pass these to VASP through \texttt{MAGMOM}. VASP then uses these magnetic moments to initialize a spin-polarized calculation before the spin density is updated self-consistently. We test this method using the charge-informed model CHGNet~\citep{deng2023chgnet}. Second, the smooth spin-difference density
$\tilde{\rho}^{-}(\mathbf r)
=
\tilde{\rho}_{\uparrow}(\mathbf r)
-
\tilde{\rho}_{\downarrow}(\mathbf r)$, can be predicted directly from atomic structure. Thus, we construct a spin-adapted version of ELECTRAFI, spin-ELECTRAFI, that reuses the Gaussian parameters of the valence density ELECTRAFI model to learn a second set of signed weights,
\begin{equation}
    \hat{\tilde{\rho}}^{-}(\mathbf r)
    =
    \sum_g
    w_g^{-}
    \phi_g(\mathbf r;\boldsymbol{\mu}_g,\mathbf{\Sigma}_g).
    \label{eq:spin-electrafi}
\end{equation}
We similarly construct spin-AugNet using the same equivariant architecture to
predict the spin-difference PAW augmentation occupancies
$\mathbf d_a^{-}=\{d^{LM,-}_{a,ij}\}_{ijLM}$ directly, using a zero reference
rather than the free-atom SAD reference. However, the net spin moment provides a global constraint on the spin-difference density, so a third hybrid option factorizes the predicted smooth spin density into a global magnetic state and a normalized spatial distribution,
\begin{equation}
    \hat q_\theta(\mathbf r\mid X)
    =
    \frac{
        \hat{\tilde{\rho}}^{-}_{\mathrm{raw}}(\mathbf r\mid X)
    }{
        \int_{\Omega}
        \hat{\tilde{\rho}}^{-}_{\mathrm{raw}}(\mathbf r\mid X)\,dV
    },
    \qquad
    \hat{\tilde{\rho}}^{-}(\mathbf r\mid X)
    =
    \hat M_{\mathrm{CHGNet}}(X)\,
    \hat q_\theta(\mathbf r\mid X).
    \label{eq:spin-moment-constraint}
\end{equation}
We can therefore use CHGNet to model the global magnetic state and constrain the high-dimensional spatial distribution predicted by spin-ELECTRAFI.
During training, we use the ground-truth magnetic moment $M_{\mathrm{DFT}}$ to constrain the spin density, and replace it with $\hat M_{\mathrm{CHGNet}}(X)$ at inference. The model is trained jointly on $\tilde{\rho}^{+}$ and
$\tilde{\rho}^{-}$ through the combined loss $\mathcal L
    =
    \mathcal L_{\tilde{\rho}^{+}}
    +
    \lambda_{\mathrm{spin}}
    \mathcal L_{\tilde{\rho}^{-}}$, where $\lambda_{\mathrm{spin}}$ is a hyperparameter. The loss is adapted to magnetic and non-magnetic structures as detailed in Appendix~\ref{app:magnetic-init}. We compare all three approaches in Section~\ref{sec:experiments}.
\begin{table}[t!]
\centering
\small
\caption{
Component ablation for PAW DFT initialization.
The matrix specifies the electronic components in each experiment,
with SCF step savings relative to the default SAD initialization reported
separately for non-magnetic and magnetic Materials Project structures.
\scmark{} denotes converged (Oracle) initialization, \sxmark \ denotes SAD/default
initialization. \textsuperscript{\textbf{\textcolor{blue}{mm}}} denotes spin initialization using atomic magnetic moments.
}
\setlength{\tabcolsep}{0pt}
\renewcommand{\arraystretch}{1.08}

\begin{tabular}{
    @{}
    >{\raggedleft\arraybackslash}m{0.18\linewidth}
    *{9}{C{0.0903333\linewidth}}
    @{}
}

&
\shortstack{Spin\\channel} &
\shortstack{Aug.\\only} &
\shortstack{Valence\\only} &
Default &
\shortstack{Valence +\\aug.} &
\shortstack{mm\\only} &
\shortstack{Smooth\\grids} &
\shortstack{Val. + aug.\\+ mm} &
Oracle
\\
\midrule

Valence density $\tilde{\rho}^{+}$
& \sxmark
& \sxmark
& \scmark
& \sxmark
& \scmark
& \sxmark
& \scmark
& \scmark
& \scmark
\\

Spin density $\tilde{\rho}^{-}$
& \scmark
& \sxmark
& \sxmark
& \sxmark
& \sxmark
& \mlmark
& \scmark
& \mlmark
& \scmark
\\

Valence aug. $\mathbf d^{+}$
& \sxmark
& \scmark
& \sxmark
& \sxmark
& \scmark
& \sxmark
& \sxmark
& \scmark
& \scmark
\\

Spin aug. $\mathbf d^{-}$
& \scmark
& \scmark
& \sxmark
& \sxmark
& \sxmark
& \mlmark
& \sxmark
& \mlmark
& \scmark
\\

\addlinespace[0.25em]
\midrule
\addlinespace[0.1em]

Non-mag. [\%] 
& \nega{9.7} 
& \nega{14.3}
& \pos{0.5}
& $0$
& \pos{13.5}
& \pos{12.0}
& \pos{10.2}
& \pos{47.6}
& \pos{49.0}
\\

Mag. [\%] 
& \nega{70.7}
& \nega{65.0}
& \nega{29.4}
& $0$
& \nega{11.0}
& \nega{3.1}
& \nega{4.2}
& \pos{24.6}
& \pos{55.4}
\\

\bottomrule

\end{tabular}

\vspace{0.2em}

\label{tab:scf_component_ablation}
\end{table}
\section{Experiments}
\label{sec:experiments}
The Complete Neural Electronic Initializer in Figure~\ref{fig:init_workflow} satisfies the three initialization criteria \critref{1}-\critref{3}. We now perform the experiments required to demonstrate \critref{4}-\critref{7}.
\paragraph{Component ablations (\critref{6}).}
We isolate the contribution of each PAW initialization component to SCF
convergence, using the same Materials Project~\citep{jain2013commentary} (MP)
densities and structures evaluated in~\citet{koker2024higher} and
~\citet{elsborg2026electrafi}. As shown in
Table~\ref{tab:scf_component_ablation}, the "Valence only" setting directly exposes the limitation of prior approaches: even with the converged valence density $\tilde{\rho}^{+}$, leaving augmentation and spin at their default values provides no benefit for non-magnetic structures and substantially worsens magnetic calculations. This is consistent with the conclusion of \citet{sunshine2023chemical}, who found no practical acceleration when combining an ML valence density prediction with augmentation occupancies from a zero-step VASP DFT calculation. Acceleration therefore requires PAW augmentation and spin initialization (\critref{2}-\critref{3}) for practical acceleration. The Oracle setting uses only converged quantities to set a practical upper bound on achievable acceleration: $49.0\%$ and $55.4\%$ SCF step reduction for non-magnetic and magnetic structures, respectively. Comparing valence + augmentation initialization $(\tilde{\rho}^{+},\mathbf d^{+})$ with Oracle isolates the importance of spin initialization, since adding the spin-dependent components recovers much of the remaining acceleration for both non-magnetic and magnetic structures. Full results are provided in Appendix~\ref{app:baseline_scf} and Table~\ref{table:scf_exps}.

\paragraph{Magnetic initialization.}
Table~\ref{tab:magnetic-init-selection} compares the three magnetic initialization strategies introduced in Section~\ref{sec:method}: CHGNet magnetic moments, explicit spatial initialization using spin-ELECTRAFI and spin-AugNet, and the hybrid model combining CHGNet-constrained spin-ELECTRAFI with spin-AugNet. We compare against Oracle spin channel initialization $(\tilde{\rho}^{-},\mathbf d^{-})$ and Oracle magnetic moments to isolate the acceleration available from each representation. Oracle results show that magnetic moments recover most of the available acceleration for non-magnetic structures, but substantially less for magnetic structures. Using CHGNet moments results in the same overall picture. Direct spatial initialization with spin-ELECTRAFI and spin-AugNet provides the required spin-dependent density representation, but these predictions are inaccurate and lead to less acceleration for magnetic systems, particularly when coupled with the ML valence and augmentation methods. In our hybrid model, constraining spin-ELECTRAFI with CHGNet reduces magnetic prediction error significantly, while spin-AugNet supplies the corresponding spin-difference augmentation occupancies $\mathbf d^{-}$, recovering a larger fraction of the available acceleration while retaining performance on non-magnetic structures. Our Complete Neural Electronic Initializer in Figure~\ref{fig:init_workflow} therefore uses the hybrid model. Details on the magnetic initialization models and experiments are provided in Appendix~\ref{app:magnetic-init}, with hyperparameters in~\ref{app:hyperparameters}.

\begin{table}[t!]
\centering
\caption{%
Spin-difference density accuracy and SCF step reduction for magnetic
initializations using Oracle and ML valence and augmentation on the Materials Project test set.
}
\label{tab:magnetic-init-selection}
\small
\setlength{\tabcolsep}{3.0pt}
\renewcommand{\arraystretch}{1.14}
\begin{tabularx}{\linewidth}{@{}
    >{\raggedright\arraybackslash}p{0.20\linewidth}
    >{\raggedright\arraybackslash}p{0.11\linewidth}
    >{\centering\arraybackslash}p{0.09\linewidth}
    >{\centering\arraybackslash}X
    >{\centering\arraybackslash}X
    >{\centering\arraybackslash}X
@{}}
\toprule
\textbf{Magnetic initialization} & \textbf{Subset} & \textbf{MAE}
    & \textbf{Oracle valence + augmentation}
    & \textbf{ELECTRAFI + AugNet}
    & \textbf{ChargE3Net + AugNet} \\
\midrule
\multirow{2}{*}{Oracle $(\tilde{\rho}^{-},\mathbf d^{-})$}
    & Non-mag.  & --     & 49.0\% & 23.3\% & 29.1\% \\
    & Magnetic  & --     & 55.4\% & 29.3\% & 31.9\% \\
\addlinespace[0.35em]
\multirow{2}{*}{Oracle moments}
    & Non-mag.  & 0.028  & 47.6\% & 22.6\% & 29.0\% \\
    & Magnetic  & 4.658  & 24.7\% &  6.2\% &  9.5\% \\
\addlinespace[0.15em]
\cmidrule(lr){1-6}
\addlinespace[0.05em]
\multicolumn{6}{@{}l}{\textbf{Models}} \\
\addlinespace[0.05em]
\multirow{2}{*}{CHGNet moments}
    & Non-mag.  & 0.347  & 38.7\% & 19.4\% & 25.9\% \\
    & Magnetic  & 5.662  & 20.5\% &  1.7\% &  3.8\% \\
\addlinespace[0.35em]
\multirow{2}{*}{\shortstack[l]{spin-ELECTRAFI\\+ spin-AugNet}}
    & Non-mag.  & 0.194  & 40.8\% & 18.1\% & 24.4\% \\
    & Magnetic  & 5.950  & 12.8\% & -1.0\% &  0.1\% \\
\addlinespace[0.55em]
\multirow{2}{*}{\shortstack[l]{Hybrid\\+ spin-AugNet}}
    & Non-mag.  & 0.316  & 38.8\% & 18.7\% & 23.9\% \\
    & Magnetic  & 2.870  & 24.2\% & 10.2\% & 12.0\% \\
\bottomrule
\end{tabularx}
\end{table}

\begin{figure*}[t!]
    \centering

    {
    \captionof{table}{
    PAW augmentation occupancy prediction errors. CJM is evaluated on its system-specific $\mathrm{MoS}_2$ structure. AugNet is evaluated on the MP and GNoME test data.
    }
    \label{table:augnet_accuracy_reference}

    \small
    \setlength{\tabcolsep}{7pt}

    \begin{tabular}{lccccc}
    \toprule
    \textbf{Model:}
    & CJM
    & AugNet
    & AugNet
    & spin-AugNet
    & spin-AugNet
    \\
    
    \textbf{Data:}
    & $\mathrm{MoS}_2$
    & MP
    & GNoME
    & MP
    & GNoME
    \\
    \midrule
    
    MAE $\downarrow$
    & 0.0130
    & 0.0041
    & 0.0062
    & 0.0025
    & 0.0020
    \\
    
    RMSE $\downarrow$
    & 0.0459
    & 0.0119
    & 0.0262
    & 0.0098
    & 0.0129
    \\
    
    \bottomrule
    \end{tabular}
    }



    \vspace{0.2cm}

    \includegraphics[
        width=0.99\textwidth
    ]{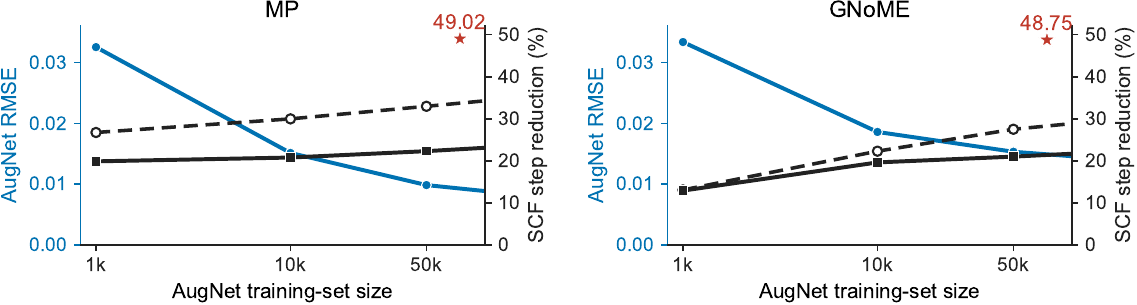}
    \par\vspace{0.1cm}
    \includegraphics[
        width=0.9\textwidth
    ]{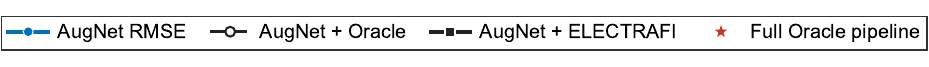}
    \par\vspace{-0.1cm}
    \caption{
    AugNet scaling with training set size on MP and OOD GNoME, using either the converged Oracle or the ELECTRAFI model for valence density initialization.
    }
    \label{fig:augnet-scaling}

\end{figure*}

\paragraph{AugNet performance (\critref{4}--\critref{5}).}
We test AugNet's ability to improve DFT initialization by training on progressively larger Materials Project subsets and evaluating the non-magnetic MP and GNoME test sets of~\citet{elsborg2026electrafi}. Figure~\ref{fig:augnet-scaling} shows that increasing the training set reduces RMSE on both datasets and improves SCF convergence. Pairing AugNet models with Oracle valence density or ELECTRAFI shows that lower augmentation error translates into better initialization with both converged and learned densities. Scaling saturates earlier on MP, while GNoME benefits from additional data. The full model reaches an augmentation MAE/RMSE of $0.0041/0.0118$ on MP and $0.0062/0.0262$ on GNoME (Table~\ref{table:augnet_accuracy_reference}). For context, CJM reports $0.0130/0.0459$ MAE/RMSE on its system-specific $\mathrm{MoS}_2$ dataset~\citep{focassio2024covariant}, and EdenGNN an MAE of $0.0085$ on non-magnetic MP structures recomputed with different VASP settings~\citep{li2025efficient}. Neither is a matched benchmark, since datasets, settings and aggregation differ, but both indicate the scale of coefficient-space errors. AugNet can also be efficiently adapted to the $\mathrm{MoS}_2$ PAW setup through fine-tuning (Appendix~\ref{app:augnet-accuracy-transfer}).  

\begin{figure*}[t!]
    \centering

{
\captionof{table}{%
End-to-end performance of our Complete Neural Electronic Initializer (CNEI), using either ELECTRAFI (EFI) or ChargE3Net (C3Net) as the valence backbone. Total time includes ML initialization and DFT execution. SCF step savings are reported relative to Default. Oracle acceleration captured (OAC) is calculated as defined in Equation~\ref{eq:oac}.
}
\label{tab:end-to-end-summary}
\small
\setlength{\tabcolsep}{8pt}
\renewcommand{\arraystretch}{1.08}
\begin{tabular*}{0.92\textwidth}{@{\extracolsep{\fill}}llcccc}
\toprule
Dataset & Metric & Default & Oracle & CNEI-EFI & CNEI-C3Net \\
\midrule
\multirow{6}{*}{\textbf{MP}}
    & NMAE [\%]      $\downarrow$         & --     & --     & 0.58           & 0.54  \\
    \addlinespace[0.15em]
    & SCF steps $\downarrow$              & 22.05  & 10.41  & 19.05          & 18.36  \\
    & SCF steps saved [\%] $\uparrow$     & --     & 52.78  & 13.62          & 16.76  \\
\addlinespace[0.15em]
    & Total time [s] $\downarrow$         & 623.84 & 302.04 & 530.04         & 585.26 \\
    & Total time saved [\%] $\uparrow$    & --     & 51.58  & 15.04          & 6.18   \\
    & OAC [\%] $\uparrow$                 & 0.0    & 100.0  & \textbf{29.15} & 11.99  \\
\addlinespace[0.10em]
\midrule
\addlinespace[0.30em]
\multirow{6}{*}{\textbf{GNoME}}
    & NMAE [\%]      $\downarrow$         & --     & --     & 0.93           & 0.69  \\
    \addlinespace[0.15em]
    & SCF steps $\downarrow$              & 16.30  & 7.89   & 11.87          & 11.45  \\
    & SCF steps saved [\%] $\uparrow$     & --     & 51.63  & 27.17          & 29.79  \\
\addlinespace[0.15em]
    & Total time [s] $\downarrow$         & 188.99 & 112.59 & 141.43         & 174.02 \\
    & Total time saved [\%] $\uparrow$    & --     & 40.43  & 25.17          & 7.92   \\
    & OAC [\%] $\uparrow$                 & 0.0    & 100.0  & \textbf{62.25} & 19.59  \\
\bottomrule
\end{tabular*}
}



            \vspace{0.32cm}
    \includegraphics[width=0.71\textwidth]{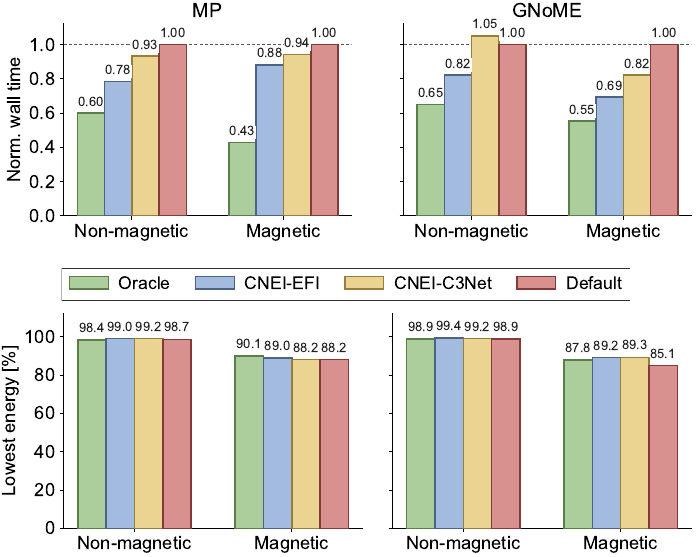}
    \par\vspace{0.49cm}
    \caption{
    \textbf{Top:} Wall-time comparison of Default, Oracle, and our Complete Neural Electronic Initializer (CNEI), using ELECTRAFI (EFI) or ChargE3Net (C3Net) for valence density predictions. Wall time is normalized relative to Default $=1.0$. Values $< 1$ indicate acceleration, while values $>1$ indicate slowdown.
    \textbf{Bottom:} The proportion of calculations for each method that are within 1 meV of the lowest energy recorded for any method.
    }
    \label{fig:finalfig}

\end{figure*}

\paragraph{Complete Neural Electronic Initializer and reference-free acceleration (\critref{7}).}
We finally evaluate the ability of our Complete Neural Electronic Initializer to provide reference-free acceleration. We use AugNet and the hybrid magnetic model from Section~\ref{sec:method}, and combine them with either ELECTRAFI~\citep{elsborg2026electrafi} or ChargE3Net~\citep{koker2024higher} as the valence density model.
Figure~\ref{fig:finalfig} shows the effect of the Complete Neural Electronic Initializer (CNEI) on total wall time for both the non-magnetic and magnetic subsets of MP and GNoME. As in ~\citet{elsborg2026electrafi}, we report the wall time reductions adjusted for ML inference time, and compare to the Default and Oracle wall time numbers in Table~\ref{tab:end-to-end-summary} across the full test sets. We include Oracle acceleration captured (OAC), which is the fraction of the reduction achieved by Oracle that is recovered by ML initialization:
\begin{equation}
\mathrm{OAC}_{\mathrm{ML}}
=
\frac{\mathrm{ML\ wall\text{-}time\ saving}\;(\%)}
{\mathrm{Oracle\ wall\text{-}time\ saving}\;(\%)}
\times 100\%.
\label{eq:oac}
\end{equation}
The initializer reduces total wall time by $15.04\%$ on MP and $25.17\%$ on GNoME using ELECTRAFI as the valence backbone (CNEI-EFI), corresponding to $\mathrm{OAC}_{\mathrm{CNEI-EFI}}(\mathrm{MP})=29.15\%$ and $\mathrm{OAC}_{\mathrm{CNEI-EFI}}(\mathrm{GNoME})=62.25 \%$. Using ChargE3Net as the valence model (CNEI-C3Net) produces larger reductions in DFT execution time, but its inference cost limits end-to-end savings to $6.18\%$ and $7.92\%$ ($\mathrm{OAC}_{\mathrm{CNEI-C3Net}}(\mathrm{MP})=11.99\%$ and $\mathrm{OAC}_{\mathrm{CNEI-C3Net}}(\mathrm{GNoME})=19.59 \%$). The AugNet and magnetic models add virtually no overhead, so the valence density model dominates inference cost. For both CNEI-EFI and CNEI-C3Net, Figure~\ref{fig:finalfig} shows that the room for improvement is largest on magnetic structures, which are not accelerated as much as non-magnetic structures. The lower panel of Figure~\ref{fig:finalfig} shows that learned initializations do not alter the converged outcome relative to either Default or Oracle initialization, with all four methods reaching the lowest observed energy at similar rates. On magnetic GNoME structures, CNEI reaches it more often than Default ($89.2\%$ against $85.1\%$), suggesting that learned spin initialization can also steer SCF toward lower-energy magnetic states. Full numerical results are in Appendix~\ref{app:end-to-end-results}.

\section{Discussion \& Limitations}
Our results show that practical PAW initialization for materials DFT acceleration is a multi-component prediction problem. Valence density alone is insufficient, and augmentation and magnetic initialization are necessary to achieve reference-free acceleration. Our Complete Neural Electronic Initializer is, to our knowledge, the first method to achieve end-to-end acceleration in spin-polarized PAW DFT without any electronic quantity from a converged target calculation.
However, better training metrics are needed. The GNoME evaluations have larger density errors, yet the wall time reduction is larger (Tables~\ref{table:augnet_accuracy_reference}-\ref{tab:end-to-end-summary} and Figure~\ref{fig:finalfig}), so current density metrics are imperfect proxies for solver performance. Solver-aware approaches such as~\citet{eberhard2026SAIL} are promising for explicitly optimizing execution time, but difficult to apply to VASP because they require access to solver gradients. Furthermore, the predicted smooth spin-difference density and spin-difference PAW augmentation occupancies are less accurate than the smooth valence density predictions, and because CHGNet predicts moment magnitudes, the hybrid model cannot distinguish antiferromagnetic from ferromagnetic order. Table~\ref{tab:scf_component_ablation} shows that improving these components is a clear route towards closing the gap to Oracle initialization. New charge-informed models for magnetic property prediction could aid this development~\citep{li2023deep,xu2025spin}. Predicted electronic quantities also depend on the material distribution and the PAW and DFT setup. Future models could explicitly encode functionals, pseudopotentials, and other DFT settings to enable broader transfer. Alternatively, as shown for AugNet in Appendix~\ref{app:augnet-accuracy-transfer}, models can be adapted to new PAW setups through transfer learning.
Figure~\ref{fig:finalfig} shows that ML initialization reaches the lowest converged energies at the same rate as default DFT initialization, so we view DFT acceleration as complementary to improving MLIPs. The learned model changes only the initialization, while the final energy is still obtained by solving the original DFT problem. This can therefore accelerate DFT calculations where surrogates are not sufficient, as well as the generation of data for increasingly accurate surrogates. The gains apply wherever a calculation starts without a converged density, such as single points on new or perturbed structures, which dominate non-equilibrium datasets like OMat24~\citep{barros2026open}, and the first ionic step of a relaxation. We expect smaller benefits for later relaxation steps, since these can already restart from the previous step's density. Our results therefore establish the components and evaluation required for complete neural electronic initialization in PAW-based periodic DFT for materials and show that complete, reference-free initialization is achievable. We hope this encourages new models for all components required for the complete initializer, since our ablations show that each one limits the acceleration that can be reached.

\newpage

\backmattersection{Software and Data}
The full codebase for the Complete Neural Electronic Initializer and its associated experiments is publicly available at:
\url{https://github.com/aerte/neural_paw_dft}, under the license specified in the repository.

The DFT calculations in this paper have been performed using the ab-initio total-energy and
molecular-dynamics program VASP (Vienna ab-initio simulation program)
developed at the Fakultät für Physik of the Universität Wien~\citep{kresse1996efficient,kresse1999ultrasoft}.

\backmattersection{Acknowledgments}
The authors acknowledge financial support from the Independent Research Foundation Denmark with grant no. 3164-00297B (ADANA), the Novo Nordisk Foundation with grant number NNF25OC0101622 (AutoMLP), and the Pioneer Center for Accelerating P2X Materials Discovery (CAPeX), DNRF grant number P3.

\bibliography{iclr2027_conference}

@article{febrer2024graph2mat,
  title={Graph2Mat: universal graph to matrix conversion for electron density prediction},
  author={Febrer, Pol and J{\o}rgensen, Peter Bj{\o}rn and Pruneda, Miguel and Garc{\'\i}a, Alberto and Ordej{\'o}n, Pablo and Bhowmik, Arghya},
  journal={Machine Learning: Science and Technology},
  volume={6},
  number={2},
  pages={025013},
  year={2025},
  publisher={IOP Publishing}
}

@article{jorgensen2022equivariant,
  title={Equivariant graph neural networks for fast electron density estimation of molecules, liquids, and solids},
  author={J{\o}rgensen, Peter Bj{\o}rn and Bhowmik, Arghya},
  journal={npj Computational Materials},
  volume={8},
  number={1},
  pages={183},
  year={2022},
  publisher={Nature Publishing Group UK London}
}

@article{fu2024recipe,
  title={A recipe for charge density prediction},
  author={Fu, Xiang and Rosen, Andrew and Bystrom, Kyle and Wang, Rui and Musaelian, Albert and Kozinsky, Boris and Smidt, Tess and Jaakkola, Tommi},
  journal={Advances in Neural Information Processing Systems},
  volume={37},
  pages={9727--9752},
  year={2024}
}

@article{batatia2022mace,
  title={MACE: Higher order equivariant message passing neural networks for fast and accurate force fields},
  author={Batatia, Ilyes and Kovacs, David P and Simm, Gregor and Ortner, Christoph and Cs{\'a}nyi, G{\'a}bor},
  journal={Advances in Neural Information Processing Systems},
  volume={35},
  pages={11423--11436},
  year={2022}
}

@article{koker2024higher,
  title={Higher-order equivariant neural networks for charge density prediction in materials},
  author={Koker, Teddy and Quigley, Keegan and Taw, Eric and Tibbetts, Kevin and Li, Lin},
  journal={npj Computational Materials},
  volume={10},
  number={1},
  pages={161},
  year={2024},
  publisher={Nature Publishing Group UK London}
}

@article{kim2024gaussian,
  title={Gaussian Plane-Wave Neural Operator for Electron Density Estimation},
  author={Kim, Seongsu and Ahn, Sungsoo},
  journal={arXiv preprint arXiv:2402.04278},
  year={2024}
}

@article{cheng2024equivariant,
  title={Equivariant neural operator learning with graphon convolution},
  author={Cheng, Chaoran and Peng, Jian},
  journal={Advances in Neural Information Processing Systems},
  volume={36},
  year={2024}
}

@article{hohenberg1964inhomogeneous,
  title={Inhomogeneous electron gas},
  author={Hohenberg, Pierre and Kohn, Walter},
  journal={Physical review},
  volume={136},
  number={3B},
  pages={B864},
  year={1964},
  publisher={APS}
}

@article{kohn1965self,
  title={Self-consistent equations including exchange and correlation effects},
  author={Kohn, Walter and Sham, Lu Jeu},
  journal={Physical review},
  volume={140},
  number={4A},
  pages={A1133},
  year={1965},
  publisher={APS}
}

@article{blochl1994projector,
  title={Projector augmented-wave method},
  author={Bl{\"o}chl, Peter E},
  journal={Physical review B},
  volume={50},
  number={24},
  pages={17953},
  year={1994},
  publisher={APS}
}

@article{elsborg2025electra,
  title={Electra: A cartesian network for 3d charge density prediction with floating orbitals},
  author={Elsborg, Jonas and Thiede, Luca and Aspuru-Guzik, Al{\'a}n and Vegge, Tejs and Bhowmik, Arghya},
  journal={Advances in Neural Information Processing Systems},
  volume={38},
  pages={28092--28121},
  year={2026}
}

@article{qu2024importance,
  title={The importance of being scalable: Improving the speed and accuracy of neural network interatomic potentials across chemical domains},
  author={Qu, Eric and Krishnapriyan, Aditi},
  journal={Advances in Neural Information Processing Systems},
  volume={37},
  pages={139030--139053},
  year={2024}
}

@article{batzner20223,
  title={E (3)-equivariant graph neural networks for data-efficient and accurate interatomic potentials},
  author={Batzner, Simon and Musaelian, Albert and Sun, Lixin and Geiger, Mario and Mailoa, Jonathan P and Kornbluth, Mordechai and Molinari, Nicola and Smidt, Tess E and Kozinsky, Boris},
  journal={Nature communications},
  volume={13},
  number={1},
  pages={2453},
  year={2022},
  publisher={Nature Publishing Group UK London}
}

@article{kresse1996efficient,
  title={Efficient iterative schemes for ab initio total-energy calculations using a plane-wave basis set},
  author={Kresse, Georg and Furthm{\"u}ller, J{\"u}rgen},
  journal={Physical review B},
  volume={54},
  number={16},
  pages={11169},
  year={1996},
  publisher={APS}
}

@article{jain2013commentary,
  title={Commentary: The Materials Project: A materials genome approach to accelerating materials innovation},
  author={Jain, Anubhav and Ong, Shyue Ping and Hautier, Geoffroy and Chen, Wei and Richards, William Davidson and Dacek, Stephen and Cholia, Shreyas and Gunter, Dan and Skinner, David and Ceder, Gerbrand and others},
  journal={APL materials},
  volume={1},
  number={1},
  year={2013},
  publisher={AIP Publishing}
}

@article{payne1992iterative,
  title={Iterative minimization techniques for ab initio total-energy calculations: molecular dynamics and conjugate gradients},
  author={Payne, Mike C and Teter, Michael P and Allan, Douglas C and Arias, TA and Joannopoulos, ad JD},
  journal={Reviews of modern physics},
  volume={64},
  number={4},
  pages={1045},
  year={1992},
  publisher={APS}
}

@article{gavini2023roadmap,
  title={Roadmap on electronic structure codes in the exascale era},
  author={Gavini, Vikram and Baroni, Stefano and Blum, Volker and Bowler, David R and Buccheri, Alexander and Chelikowsky, James R and Das, Sambit and Dawson, William and Delugas, Pietro and Dogan, Mehmet and others},
  journal={Modelling and Simulation in Materials Science and Engineering},
  volume={31},
  number={6},
  pages={063301},
  year={2023},
  publisher={IOP Publishing}
}

@article{de2023nanosecond,
  title={Nanosecond MD of battery cathode materials with electron density description},
  author={de Blasio, Paolo Vincenzo Freiesleben and Jorgensen, Peter Bj{\o}rn and Lastra, Juan Maria Garcia and Bhowmik, Arghya},
  journal={Energy Storage Materials},
  volume={63},
  pages={103023},
  year={2023},
  publisher={Elsevier}
}

@article{focassio2024covariant,
  title={Covariant Jacobi-Legendre expansion for total energy calculations within the projector augmented wave formalism},
  author={Focassio, Bruno and Domina, Michelangelo and Patil, Urvesh and Fazzio, Adalberto and Sanvito, Stefano},
  journal={Physical Review B},
  volume={110},
  number={18},
  pages={184106},
  year={2024},
  publisher={APS}
}

@article{merchant2023scaling,
  title={Scaling deep learning for materials discovery},
  author={Merchant, Amil and Batzner, Simon and Schoenholz, Samuel S and Aykol, Muratahan and Cheon, Gowoon and Cubuk, Ekin Dogus},
  journal={Nature},
  volume={624},
  number={7990},
  pages={80--85},
  year={2023},
  publisher={Nature Publishing Group UK London}
}

@article{kresse1999ultrasoft,
  title={From ultrasoft pseudopotentials to the projector augmented-wave method},
  author={Kresse, Georg and Joubert, Daniel},
  journal={Physical review b},
  volume={59},
  number={3},
  pages={1758},
  year={1999},
  publisher={APS}
}

@misc{elsborg2026electrafi,
      title={Global Plane Waves From Local Gaussians: Periodic Charge Densities in a Blink}, 
      author={Jonas Elsborg and Felix Ærtebjerg and Luca Thiede and Alán Aspuru-Guzik and Tejs Vegge and Arghya Bhowmik},
      year={2026},
      eprint={2601.19966},
      archivePrefix={arXiv},
      primaryClass={cond-mat.mtrl-sci},
      url={https://arxiv.org/abs/2601.19966}, 
}

@article{chen2022universal,
  title={A universal graph deep learning interatomic potential for the periodic table},
  author={Chen, Chi and Ong, Shyue Ping},
  journal={Nature Computational Science},
  volume={2},
  number={11},
  pages={718--728},
  year={2022},
  publisher={Nature Publishing Group US New York}
}

@article{deng2023chgnet,
  title={CHGNet as a pretrained universal neural network potential for charge-informed atomistic modelling},
  author={Deng, Bowen and Zhong, Peichen and Jun, KyuJung and Riebesell, Janosh and Han, Kevin and Bartel, Christopher J and Ceder, Gerbrand},
  journal={Nature Machine Intelligence},
  volume={5},
  number={9},
  pages={1031--1041},
  year={2023},
  publisher={Nature Publishing Group UK London}
}

@article{batatia2025foundation,
  title={A foundation model for atomistic materials chemistry},
  author={Batatia, Ilyes and Benner, Philipp and Chiang, Yuan and Elena, Alin M and Kov{\'a}cs, D{\'a}vid P and Riebesell, Janosh and Advincula, Xavier R and Asta, Mark and Avaylon, Matthew and Baldwin, William J and others},
  journal={The Journal of chemical physics},
  volume={163},
  number={18},
  year={2025},
  publisher={AIP Publishing}
}

@article{neumann2024orb,
  title={Orb: A fast, scalable neural network potential},
  author={Neumann, Mark and Gin, James and Rhodes, Benjamin and Bennett, Steven and Li, Zhiyi and Choubisa, Hitarth and Hussey, Arthur and Godwin, Jonathan},
  journal={arXiv preprint arXiv:2410.22570},
  year={2024}
}

@inproceedings{klockowfunction,
  title={A Function-Centric Graph Neural Network Approach for Predicting Electron Densities},
  author={Klockow, Manuel V and Ickler, Marc K and Lippmann, Peter and Hamprecht, Fred A},
  booktitle={The Fourteenth International Conference on Learning Representations},
  year={2026}
}

@article{calcaterra2008approximating,
  title={Approximating with gaussians},
  author={Calcaterra, Craig and Boldt, Axel},
  journal={arXiv preprint arXiv:0805.3795},
  year={2008}
}

@misc{eberhard2026SAIL,
      title={Transferable SCF-Acceleration through Solver-Aligned Initialization Learning}, 
      author={Eike S. Eberhard and Viktor Kotsev and Timm Güthle and Stephan Günnemann},
      year={2026},
      eprint={2604.21657},
      archivePrefix={arXiv},
      primaryClass={cs.LG},
      url={https://arxiv.org/abs/2604.21657}, 
}

@article{pymatgen,
title = {Python Materials Genomics (pymatgen): A robust, open-source python library for materials analysis},
journal = {Computational Materials Science},
volume = {68},
pages = {314-319},
year = {2013},
issn = {0927-0256},
doi = {https://doi.org/10.1016/j.commatsci.2012.10.028},
url = {https://www.sciencedirect.com/science/article/pii/S0927025612006295},
author = {Shyue Ping Ong and William Davidson Richards and Anubhav Jain and Geoffroy Hautier and Michael Kocher and Shreyas Cholia and Dan Gunter and Vincent L. Chevrier and Kristin A. Persson and Gerbrand Ceder}
}

@article{riebesell2025framework,
  title={A framework to evaluate machine learning crystal stability predictions},
  author={Riebesell, Janosh and Goodall, Rhys EA and Benner, Philipp and Chiang, Yuan and Deng, Bowen and Ceder, Gerbrand and Asta, Mark and Lee, Alpha A and Jain, Anubhav and Persson, Kristin A},
  journal={Nature Machine Intelligence},
  volume={7},
  number={6},
  pages={836--847},
  year={2025},
  publisher={Nature Publishing Group UK London}
}

@article{li2023deep,
  title={Deep-learning electronic-structure calculation of magnetic superstructures},
  author={Li, He and Tang, Zechen and Gong, Xiaoxun and Zou, Nianlong and Duan, Wenhui and Xu, Yong},
  journal={Nature Computational Science},
  volume={3},
  pages={321--327},
  year={2023},
  doi={10.1038/s43588-023-00424-3}
}

@article{xu2025spin,
  title={Spin-informed universal graph neural networks for simulating magnetic ordering},
  author={Xu, Wenbin and Sanspeur, Rohan Yuri and Kolluru, Adeesh and Deng, Bowen and Harrington, Peter and Farrell, Steven and Reuter, Karsten and Kitchin, John R.},
  journal={Proceedings of the National Academy of Sciences},
  volume={122},
  number={27},
  pages={e2422973122},
  year={2025},
  doi={10.1073/pnas.2422973122}
}

@article{Choudhary_2021,
   title={Atomistic Line Graph Neural Network for improved materials property predictions},
   volume={7},
   ISSN={2057-3960},
   url={http://dx.doi.org/10.1038/s41524-021-00650-1},
   DOI={10.1038/s41524-021-00650-1},
   number={1},
   journal={npj Computational Materials},
   publisher={Springer Science and Business Media LLC},
   author={Choudhary, Kamal and DeCost, Brian},
   year={2021},
   month=Nov }

@misc{wood2026umafamilyuniversalmodels,
      title={UMA: A Family of Universal Models for Atoms}, 
      author={Brandon M. Wood and Misko Dzamba and Xiang Fu and Meng Gao and Muhammed Shuaibi and Luis Barroso-Luque and Kareem Abdelmaqsoud and Vahe Gharakhanyan and John R. Kitchin and Daniel S. Levine and Kyle Michel and Anuroop Sriram and Taco Cohen and Abhishek Das and Ammar Rizvi and Sushree Jagriti Sahoo and Zachary W. Ulissi and C. Lawrence Zitnick},
      year={2026},
      eprint={2506.23971},
      archivePrefix={arXiv},
      primaryClass={cs.LG},
      url={https://arxiv.org/abs/2506.23971}, 
}

@article{barros2026open,
  title={The Open Materials 2024 (OMat24) inorganic materials dataset and models},
  author={Barros-Luque, Luis and Shuaibi, Muhammed and Fu, Xiang and Wood, Brandon M and Dzamba, Misko and Gao, Meng and Rizvi, Ammar and Uyttendaele, Matt and Zitnick, C Lawrence and Ulissi, Zachary W},
  journal={Nature Computational Science},
  pages={1--11},
  year={2026},
  publisher={Nature Publishing Group US New York}
}

@article{kirklin2015open,
  title={The Open Quantum Materials Database (OQMD): assessing the accuracy of DFT formation energies},
  author={Kirklin, Scott and Saal, James E and Meredig, Bryce and Thompson, Alex and Doak, Jeff W and Aykol, Muratahan and R{\"u}hl, Stephan and Wolverton, Chris},
  journal={npj Computational Materials},
  volume={1},
  number={1},
  pages={15010},
  year={2015},
  publisher={Nature Publishing Group}
}

@article{curtarolo2012aflowlib,
  title={AFLOWLIB. ORG: A distributed materials properties repository from high-throughput ab initio calculations},
  author={Curtarolo, Stefano and Setyawan, Wahyu and Wang, Shidong and Xue, Junkai and Yang, Kesong and Taylor, Richard H and Nelson, Lance J and Hart, Gus LW and Sanvito, Stefano and Buongiorno-Nardelli, Marco and others},
  journal={Computational Materials Science},
  volume={58},
  pages={227--235},
  year={2012},
  publisher={Elsevier}
}

@article{choudhary2020joint,
  title={The joint automated repository for various integrated simulations (JARVIS) for data-driven materials design},
  author={Choudhary, Kamal and Garrity, Kevin F and Reid, Andrew CE and DeCost, Brian and Biacchi, Adam J and Hight Walker, Angela R and Trautt, Zachary and Hattrick-Simpers, Jason and Kusne, A Gilad and Centrone, Andrea and others},
  journal={npj computational materials},
  volume={6},
  number={1},
  pages={173},
  year={2020},
  publisher={Nature Publishing Group UK London}
}

@article{kim2026high,
  title={High-order equivariant flow matching for density functional theory hamiltonian prediction},
  author={Kim, Seongsu and Kim, Nayoung and Kim, Dongwoo and Ahn, Sungsoo},
  journal={Advances in Neural Information Processing Systems},
  volume={38},
  pages={13265--13307},
  year={2026}
}

@article{kim2026machine,
  title={Machine Learning Hamiltonians are Accurate Energy-Force Predictors},
  author={Kim, Seongsu and Lee, Chanhui and Kim, Yoonho and Yun, Seongjun and Kim, Honghui and Kim, Nayoung and Park, Changyoung and Han, Sehui and Lim, Sungbin and Ahn, Sungsoo},
  journal={arXiv preprint arXiv:2602.16897},
  year={2026}
}

@article{kaniselvan2025learning,
  title={Learning from the electronic structure of molecules across the periodic table},
  author={Kaniselvan, Manasa and Miller, Benjamin Kurt and Gao, Meng and Nam, Juno and Levine, Daniel S},
  journal={arXiv preprint arXiv:2510.00224},
  year={2025}
}

@article{qin2026eac,
  title={EAC-Net: Predicting real-space charge density via equivariant atomic contributions},
  author={Qin, Xuejian and Lv, Taoyuze and Zhong, Zhicheng},
  journal={Journal of Chemical Theory and Computation},
  volume={22},
  number={9},
  pages={4813--4821},
  year={2026},
  publisher={ACS Publications}
}

@article{song2026neural,
  title={Neural network self-consistent fields for density functional theory},
  author={Song, Feitong and Feng, Ji},
  journal={npj Computational Materials},
  year={2026},
  publisher={Nature Publishing Group UK London}
}

@article{zhang2026parsec,
  title={PARSEC. py: A Python-Based Real-Space Kohn--Sham Density Functional Theory Code Accelerated by Machine Learned Charge Density},
  author={Zhang, Zeyi and Perez, Carlos Mora and Kwon, Patrick and Head-Gordon, Martin and Qian, Jin},
  journal={Journal of Computational Chemistry},
  volume={47},
  number={23},
  pages={e70482},
  year={2026},
  publisher={Wiley Online Library}
}

@article{sunshine2023chemical,
  title={Chemical properties from graph neural network-predicted electron densities},
  author={Sunshine, Ethan M and Shuaibi, Muhammed and Ulissi, Zachary W and Kitchin, John R},
  journal={The Journal of Physical Chemistry C},
  volume={127},
  number={48},
  pages={23459--23466},
  year={2023},
  publisher={ACS Publications}
}

@article{li2025efficient,
  title={Efficient E (3)-equivariant framework for universal charge density prediction},
  author={Li, Xiwen and Xin, Zaizhou and Yu, Hongyu and Zhong, Yang and Gong, Xingao and Xiang, Hongjun},
  journal={arXiv preprint arXiv:2510.00788},
  year={2025}
}
\bibliographystyle{iclr2027_conference}

\newpage
\appendix

\startcontents[appendix]

\section*{Appendix}
\subsection*{Table of Contents}

\begin{small}
\printcontents[appendix]{}{1}{\setcounter{tocdepth}{2}}
\end{small}

\clearpage

\section{Charge density models}\label{app:density_models}
\subsection{Prior work}
Machine learning charge density models learn a map from an atomic
structure $\mathcal X=\{(Z_i,\mathbf R_i)\}_{i=1}^N$ to the smooth valence
density $\tilde{\rho}^{+}(\mathbf r)$. They differ
primarily in whether the spatial dependence is represented implicitly or explicitly. Probe-based models evaluate the density by conditioning a neural network directly on each query point,
\begin{equation}
    \hat{\tilde{\rho}}^{+}(\mathbf r)
    =
    f_\theta(\mathbf r;\mathcal X),
\end{equation}
whereas basis-based models first predict coefficients or basis parameters and then evaluate an explicit expansion,
\begin{equation}
    \hat{\tilde{\rho}}^{+}(\mathbf r)
    =
    \sum_k c_k(\mathcal X)\,\phi_k(\mathbf r;\mathcal X).
\end{equation}
This distinction is architectural rather than fundamental, since probe models simply use an implicit, query-dependent basis, while basis models make the spatial representation explicit. Different approaches such as InfGCN similarly learn maps from atomic structure to continuous smooth valence density fields, but can be viewed as basis/operator variants of the same underlying problem \citep{cheng2024equivariant}.

DeepDFT introduced the probe-based formulation\citep{jorgensen2022equivariant}, and ChargE3Net replaces the DeepDFT backbone with a higher-order $E(3)$-equivariant architecture to better capture angular structure in periodic materials \citep{koker2024higher}. These approaches are flexible since they avoid choosing an explicit basis, but they are computationally intensive since evaluating a full density grid requires many query point evaluations. 

SCDP uses a spherical Gaussian basis centered on both atoms and equivariantly placed virtual centers,
\begin{equation}
    \hat{\tilde{\rho}}^{+}(\mathbf r)
    =
    \sum_{a\in\mathcal A\cup\mathcal V}
    \sum_{j\ell m}
    c_{aj\ell m}\,
    \Phi_{\alpha_{aj},\ell,m,\mathbf R_a}(\mathbf r).
\end{equation}
where $\mathcal A$ denotes atoms and $\mathcal V$ virtual centers \citep{fu2024recipe}. SCDP shows that non-atom-centered orbital bases can lead to higher accuracy, and \citet{klockowfunction} achieved a similar result in BOA by representing the density through products of atom-centered basis functions,
\begin{equation}
    \hat{\tilde{\rho}}^{+}(\mathbf r)
    =
    \sum_{(a,b)}
    \sum_{\mu,\nu}
    \Gamma_{ab\mu\nu}\,
    \omega_\mu^{Z_a}(\mathbf r-\mathbf R_a)
    \omega_\nu^{Z_b}(\mathbf r-\mathbf R_b).
\end{equation}
which resembles a local density matrix expansion and naturally places density between atoms. \\
ELECTRA takes a different explicit representation approach by replacing spherical-harmonic orbital expansions with a mixture of anisotropic floating 3D Gaussians \citep{elsborg2025electra},
\begin{equation}
    \hat{\tilde{\rho}}^{+}(\mathbf r)
    =
    \sum_{j=1}^{N_G}
    w_j\,
    \mathcal N(\mathbf r;\boldsymbol{\mu}_j,\boldsymbol{\Sigma}_j).
    \label{eq:electraeq}
\end{equation}
Each component has a signed weight $w_j$, a learned center $\boldsymbol{\mu}_j$, and a
positive-definite covariance $\boldsymbol{\Sigma}_j$. 
This ansatz uses the fact that Gaussian mixtures are universal approximators of smooth densities\citep{calcaterra2008approximating}. Compared with atom- or bond-centered orbital expansions, the basis function positions are not fixed, and are instead predicted as displacements from atoms,
\begin{equation}
    \boldsymbol{\mu}_j=\mathbf R_{a(j)}+\mathbf d_j,
\end{equation}
This removes the need for high-order spherical harmonics, resulting in inference speeds that are orders of magnitude faster than prior models. \\

\paragraph{Periodic and reciprocal-space models.}
For periodic materials, the physically natural objective is to represent densities in a way that mirrors plane-wave DFT, where periodic scalar fields are represented through reciprocal-space coefficients. \citet{kim2024gaussian} explore this by adding a plane-wave branch to a Gaussian type orbital (GTO) model, i.e., \begin{equation}
    \hat{\tilde{\rho}}^{+}(\mathbf r)
=
\hat{\tilde{\rho}}^{+}_{\mathrm{GTO}}(\mathbf r)
+
\hat{\tilde{\rho}}^{+}_{\mathrm{PW}}(\mathbf r),
\end{equation}
but this yields only modest gains compared to the GTO-only model, and performs poorly on its own.

However, ELECTRAFI showed that it is possible to extend the floating Gaussian representation of ELECTRA to periodic materials by making reciprocal space the central construction\citep{elsborg2026electrafi}. 
ELECTRAFI predicts an auxiliary non-periodic representation of $\tilde{\rho}^{+}$ similar to Equation~\ref{eq:electraeq}, and then exploits the closed-form analytical Fourier transform of each Gaussian to obtain plane-wave coefficients,
\begin{equation}
    \hat{\tilde{\rho}}^{+}(\mathbf G)
    =
    \sum_{j=1}^{N_G}
    w_j
    \exp\!\left(
        -\frac{1}{2}\mathbf G^\top \boldsymbol{\Sigma}_j \mathbf G
    \right)
    e^{-i\mathbf G\cdot\boldsymbol{\mu}_j}.
\end{equation}
The periodic real-space density is then recovered with a single inverse FFT,
\begin{equation}
    \hat{\tilde{\rho}}^{+}(\mathbf r)
    =
    \mathrm{IFFT}\!\left[
        \hat{\tilde{\rho}}^{+}(\mathbf G)
    \right](\mathbf r).
\end{equation}
Since periodicity and global Fourier structure are imposed analytically through the Poisson summation formula, ELECTRAFI's representation is the one most closely aligned with plane-wave DFT. While ChargE3Net has sufficient flexibility to achieve competitive accuracy on periodic materials, its inference times are comparable in magnitude to the DFT calculation time itself when using standard functionals \citep{elsborg2026electrafi}. The main reasons are the dense neural evaluation of every real-space grid point and the explicit summation over periodic images of the unit cell. ELECTRAFI's construction avoids both of these and achieves drastically faster inference, which also translates into end-to-end acceleration of DFT workflows, albeit still using converged properties from the reference data.

\paragraph{Other related work.}
A complementary line of work predicts electronic quantities in localized orbital representations. Graph2Mat predicts equivariant density matrices directly from atomic structure~\citep{febrer2024graph2mat}, while QHFlow and QHFlow2 learn Kohn-Sham Hamiltonians~\citep{kim2026high,kim2026machine}. QHFlow also demonstrates SCF acceleration by using the predicted Hamiltonian directly to initialize a DFT calculation, and HELM similarly targets Hamiltonian prediction across broader chemical and basis set spaces~\citep{kaniselvan2025learning}. These methods use Hamiltonian or density matrices directly in DFT frameworks formulated in localized orbital bases. NeuralSCF~\citep{song2026neural} instead learns the Kohn-Sham density map itself and iterates the learned map to self-consistency. PARSEC.py~\citep{zhang2026parsec} uses ML-predicted charge densities to initialize self-consistent Kohn-Sham calculations in a real-space finite-difference pseudopotential framework. In this representation, the predicted density can be supplied directly on the real-space grid, whereas plane-wave PAW codes such as VASP require a smooth density together with the corresponding structure-dependent PAW augmentation and, for spin-polarized calculations, spin-dependent components. These approaches therefore address related ways of reducing the cost of self-consistent electronic structure calculations using other frameworks than the periodic PAW representation considered here.
\subsection{Limitations of prior work and models}
\label{app:limitations}

Prior charge density models have demonstrated that an accurate prediction of the
smooth valence density can reduce the number of SCF iterations required by
VASP~\citep{jorgensen2022equivariant,koker2024higher,elsborg2025electra,elsborg2026electrafi}.
However, these experiments do not constitute complete initialization of a new
PAW calculation from atomic structure alone, since they are reference-dependent for augmentation and spin quantities.

For clarity, the spin-summed and spin-difference valence densities can be written as
\begin{equation}
    \rho^{+}(\mathbf r)
    =
    \rho_{\uparrow}(\mathbf r)
    +
    \rho_{\downarrow}(\mathbf r),
    \qquad
    \rho^{-}(\mathbf r)
    =
    \rho_{\uparrow}(\mathbf r)
    -
    \rho_{\downarrow}(\mathbf r).
\end{equation}
For each channel, the PAW decomposition has the form
\begin{equation}
    \rho^{\pm}(\mathbf r)
    =
    \tilde{\rho}^{\pm}(\mathbf r)
    +
    \sum_a
    \left[
        \rho^{a,\pm}(\mathbf r)
        -
        \tilde{\rho}^{a,\pm}(\mathbf r)
    \right],
    \label{eq:app-paw-spin-decomp}
\end{equation}
where $\tilde{\rho}^{\pm}$ is the smooth plane-wave component and the second
term is determined by the corresponding PAW augmentation occupancies.

\paragraph{Converged augmentation in prior SCF experiments.}
Existing generalized charge density models used for SCF acceleration predict only the smooth
spin-summed valence density $\tilde{\rho}^{+}$, following the SCF acceleration protocol initially introduced by~\citet{jorgensen2022equivariant} and subsequently adopted by later work. The initial density is therefore effectively
\begin{equation}
    \rho_{\mathrm{init}}^{+}(\mathbf r)
    =
    \underbrace{
        \tilde{\rho}_{\mathrm{ML}}^{+}(\mathbf r)
    }_{\text{predicted from structure}}
    +
    \underbrace{
        \sum_a
        \left[
            \rho_{\mathrm{test}}^{a,+}(\mathbf r)
            -
            \tilde{\rho}_{\mathrm{test}}^{a,+}(\mathbf r)
        \right]
    }_{\text{converged augmentation from the same test structure}} .
    \label{eq:app-prior-total-init}
\end{equation}
Thus, although the smooth density is predicted, the PAW augmentation
occupancies are not. They are taken from the already converged reference DFT
calculation of the exact structure whose subsequent SCF acceleration is being
measured. These quantities are therefore unavailable when DFT is run from
scratch on a genuinely new structure.

This distinction is important because augmentation is not a fixed quantity
that can simply be obtained from the PAW dataset. The dataset specifies the
partial waves, projectors, and allowed angular channels, whereas the
augmentation occupancies depend on the converged electronic state of the
material. Transferring them from the reference calculation
provides target-specific electronic information beyond the ML-predicted
valence density.

A reference-free approximation is not necessarily sufficient either.
\citet{sunshine2023chemical} obtained PAW augmentation occupancies from a VASP
calculation with zero electronic minimization steps and combined them with an
ML-predicted valence density, but found no acceleration over VASP's default
initialization and concluded that the approach had no practical value at
the time. They identified augmentation and wavefunction initialization as
remaining bottlenecks, consistent with our ablations showing that even a
converged valence density provides little acceleration when the remaining PAW
components are poorly initialized.

The limitation for spin is different. Prior generalized charge density models
do not predict $\tilde{\rho}^{-}$, nor do they predict the corresponding
spin-difference PAW augmentation occupancies. Consequently, prior SCF-acceleration
studies do not test ML initialization for general magnetic structures.
Instead, their acceleration experiments are restricted to structures classified
as non-magnetic.

However, this restriction does not eliminate the spin-dependent electronic
state as long as the underlying reference calculations are performed using default spin-polarized settings (\texttt{ISPIN=2} in VASP). A structure can have a small net magnetic
moment while still possessing a nonzero converged spin-difference density.
For such calculations, the spin channel inherited from the reference
calculation can be written schematically as
\begin{equation}
    \rho_{\mathrm{init}}^{-}(\mathbf r)
    =
    \underbrace{
        \tilde{\rho}_{\mathrm{test}}^{-}(\mathbf r)
    }_{\text{converged spin density}}
    +
    \underbrace{
        \sum_a
        \left[
            \rho_{\mathrm{test}}^{a,-}(\mathbf r)
            -
            \tilde{\rho}_{\mathrm{test}}^{a,-}(\mathbf r)
        \right]
    }_{\text{converged spin augmentation}},
    \label{eq:app-prior-spin-init}
\end{equation}
where both terms are taken from the converged DFT solution of the same test
structure rather than from an ML prediction.

Prior work does not use a dedicated spin density model for
initialization experiments, which therefore effectively prohibits general spin-polarized calculations.  Magnetic structures are not evaluated as a general reference-free acceleration problem, while even the nominally non-magnetic test calculations can retain converged spin-dependent information. An additional downside to this absence is that magnetic calculations offer the highest potential for acceleration, as we have demonstrated in this work (see Table \ref{tab:scf_component_ablation}).

\paragraph{What is required for reference-free initialization.}
A genuinely reference-free PAW initializer must instead construct all
structure-dependent electronic components without access to a converged
calculation of the test structure. In the collinear setting considered here,
this requires
\begin{equation}
    \underbrace{\tilde{\rho}_{\mathrm{ML}}^{+}}_{\text{valence density}}
    ,\qquad
    \underbrace{\{\mathbf d_{a,\mathrm{ML}}^{+}\}_a}_{\text{augmentation}}
    ,\qquad
    \underbrace{\tilde{\rho}_{\mathrm{ML}}^{-}}_{\text{spin density}}
    ,\qquad
    \underbrace{\{\mathbf d_{a,\mathrm{ML}}^{-}\}_a}_{\text{spin augmentation}},
    \label{eq:app-complete-init}
\end{equation}
all obtained from the atomic structure and quantities available before the DFT
calculation begins. Here, $\mathbf d^{+}$ and $\mathbf d^{-}$ denote the spin-summed
and spin-difference PAW augmentation occupancies, respectively.

We directly isolate these dependencies by varying which electronic components
are supplied at initialization and evaluating the resulting VASP convergence
on the Materials Project test structures used in~\citet{elsborg2026electrafi}.
Table~\ref{table:scf_exps} reports the full results of these component ablations, showing that:
\begin{itemize}
    \item
    \textbf{The previously reported benefit of valence density prediction
    depends strongly on converged PAW augmentation.}
    When the converged augmentation occupancies are removed, much of the SCF
    acceleration attributed to the predicted smooth density disappears or can
    reverse. Accurate valence density prediction alone is therefore
    insufficient for practical PAW acceleration.

    \item
    \textbf{Spin initialization constitutes a second, independent
    requirement.}
    Even nominally non-magnetic \texttt{ISPIN=2} structures benefit from
    initialization of their spin-dependent electronic state, while magnetic
    structures show an even larger dependence on accurate spin initialization.
    Prior charge density models do not address this problem and consequently
    do not establish acceleration for general magnetic materials.

    \item
    \textbf{Complete electronic initialization exposes substantially larger
    acceleration potential.}
    When the valence, augmentation, and spin-dependent components are all
    initialized accurately, the number of SCF iterations can be reduced far
    beyond what is achievable from valence density prediction alone. This
    motivates learning the previously missing augmentation and spin
    components directly from structure.
\end{itemize}

These observations identify the two missing modeling problems addressed in this work. AugNet predicts the spin-summed and spin-difference PAW augmentation occupancies, removing the need to transfer converged augmentation information from the test calculation. Separately, our charge informed spin density model uses CHGNet magnetic moment predictions to constrain ELECTRAFI and directly predicts the smooth spin-difference density $\tilde{\rho}^{-}$ required for spin-polarized initialization. Together with a valence density model, these components make it possible to initialize all structure-dependent electronic quantities from the atomic structure alone.
\clearpage
\section{VASP SCF Experiments}\label{app:baseline_scf}

\subsection{VASP Settings}\label{app:baseline_scf:vasp_settings}

All SCF calculations in this work are performed as single-point (static) calculations and differ from the corresponding default VASP calculations only in the choice of initial electron density, since we use densities predicted by machine learning models instead of the default superposition of atomic densities (SAD). To ensure a controlled comparison, we retain the parameters of the reference calculations and modify only the tags required for density initialization and output formatting. All calculations are performed with VASP 5.4.4 using the same legacy PBE PAW datasets employed in the original Materials Project calculations.




\paragraph{Materials Project.}
For every \texttt{mp-} identifier we retrieve the exact task document that produced the reference charge density via the Materials Project API and save its \texttt{POSCAR}, \texttt{INCAR}, \texttt{KPOINTS} and \texttt{POTCAR}. The structure comes from \texttt{input.structure}, the $k$-mesh from \texttt{input.kpoints}, and the pseudopotentials are reconstructed from \texttt{input.potcar\_spec} so that the POTCAR titles match the reference run element for element. Consequently, the plane-wave cutoff, the exchange-correlation functional and Hubbard-$U$ set, the smearing scheme, the electronic convergence criterion, the $k$-point mesh, and the projector set are exactly the Materials Project values for that material. 



\paragraph{GNoME.}
For GNoME calculations we use the same calculation scheme as \citet{koker2024higher, elsborg2026electrafi}, using the \texttt{pymatgen} \citep{pymatgen} \texttt{MPStatic} parameter set. We further apply the same DFT+$U$ \texttt{LMAXMIX} treatment that \citet{koker2024higher} applied.  Specifically, if DFT+$U$ is used and there are any $f$-orbitals (\texttt{LDAUL}$=3$) in the system, we set \texttt{LMAXMIX=6}, and if there are $d$-orbitals (\texttt{LDAUL}$=2$) we set \texttt{LMAXMIX=4} (excluding the case of present $f$-electrons).



\paragraph{Experiments.}
For various experiments we override the INCAR to match our experimental goal. The following list summarizes the basic parameters:

\begin{itemize}
    \item \texttt{ICHARG}$=2$ is the atomic superposition (SAD) baseline; \texttt{ICHARG}$=1$ reads the seed density from \texttt{CHGCAR}.
    \item \texttt{ISTART=0} starts the wavefunctions from scratch
    \item \texttt{LCHARG=True} generates the CHGCARs upon completion
    \item \texttt{NPAR}, \texttt{NCORE}, \texttt{KPAR}, \texttt{NSIM} are parallelization parameters that are removed, defaulting the calculation to simply use all the cores of the specified CPU. It also ensures that all calculations are given the same resources.
\end{itemize}

The spin-restricted and magnetic moment experiments require further specifications which are listed below:

\begin{itemize}
  \item \textbf{\texttt{ISPIN}$=1$ controls.} The spin channel is removed entirely: \texttt{ISPIN} is set to $1$ and the spin-only tags \texttt{MAGMOM} and \texttt{NUPDOWN} are dropped so VASP never consults them. The seed CHGCAR is correspondingly rebuilt with only the spin-summed ($+$) channel.
  
  \item \textbf{Uniform \texttt{MAGMOM} seeds.} With \texttt{ISPIN}$=2$ and a seed carrying the converged $+$-channel components $(\tilde{\rho}^{+},\mathbf d_a^{+})$ but no spin-difference channel, VASP builds the initial magnetization from \texttt{MAGMOM}. We replace MP'sper-species values with a single uniform value $m$ on every atom to limit the steps a spin-unrestricted calculation needs for non-magnetic materials. \texttt{NUPDOWN} is left as MP set it. 
  
  \item \textbf{Spin-mixing tags.} \texttt{NUPDOWN}, \texttt{AMIX\_MAG} and \texttt{BMIX\_MAG} can be overridden per run to test whether the residual spin channel cost is a mixing problem; these overrides are applied last and are otherwise inactive.
\end{itemize}

We note that if the augmentation occupancies are not formatted correctly for the given VASP version, then VASP will silently default to SAD initializations derived from the atom types and \texttt{MAGMOM}, removing any benefits gained from a better initial guess..

\paragraph{Filtering.}
Like previous works \citep{elsborg2026electrafi}, we use the magnetization filter specified by \citep{koker2024higher} to distinguish between magnetic and non-magnetic structures. The definition of the magnetic label is an absolute total magnetic moment of below $0.1 \mu_B$ and that all atomic absolute magnetic moments are below $0.1\mu_B$. For GNoME, we additionally filter out 60 structures that contain Yb, an element not present in MP dataset and therefore without a trained augmentation occupancy model. Additionally, excluding structures with convergence problems results in 951 final structures from the MP-Full test set and 1245 from GNoME.

\paragraph{Normalization.}
VASP smooth valence density grids $\tilde{\rho}^{+}$ are always normalized to the
number of valence electrons defined by the PAW dataset. As this is a predictable property that can help convergence, we normalize the ChargE3Net input predictions. A caveat is that ChargE3Net predictions is already capable of capturing the total charge within $0.1\%$ accuracy (measured on the MP non-magnetic testset). With this method, we achieve 0.05 saved steps on MP on average, i.e., a negligible difference, which was also observed by \citep{elsborg2026electrafi}.

\subsection{Magnetic Moment Initialization}\label{app:baseline_scf:magmom}

In this work, we define a structure to be non-magnetic if the absolute total magnetic moment $|\mu_B|<0.1$ \textit{and} each individual magnetic moment $|\mu_{i, B}|<0.1$. As discussed  further in appendix \ref{app:baseline_scf:baseline}, DFT calculations initialized with no spin-difference components take longer
to converge even on non-magnetic structures, despite having converged
$+$-channel components $(\tilde{\rho}^{+},\mathbf d_a^{+})$. There are three ways of providing a spin configuration guess, either setting atomic magnetic moments that are expanded into a real space SAD guess by the DFT code, or by directly initializing the smooth spin difference density
$\tilde{\rho}^{-}$ and its PAW augmentation occupancies $\mathbf d_a^{-}$. While both can be modeled with machine learning, they pose significant challenges, respectively, with the latter remaining an open challenge in the field and outside the scope of this work.

While workflows differ between DFT codes, magnetic moments are typically initialized based on some upfront observations of the structure such as the element type and the oxidation state. For atoms deemed magnetic, the initial moment is set very high to elucidate good convergence behavior, whereas non-magnetic atoms are initialized closer to 0. This way, a calculation search the potential energy surface by decreasing the magnetic moment rather than increasing it, a much harder task. As a standard of the field, the Materials Project workflow first performs two DFT relaxations with the \texttt{MPRelaxSet} (as given in \texttt{pymatgen}), shifting atom positions into more favorable positions and optimizing toward an initial guess for the spin state. Afterwards, a static DFT calculation is performed with \texttt{MPStaticSet} that determines the energy. This process has proven to be robust and results in well behaved energies and magnetic moments during data generation. However, relaxing a structure twice this way also costs more HPC resources.

\subsection{CHGNet Initialization}\label{app:baseline_scf:chgnet}

Instead of relaxing twice with DFT, practitioners could also employ one of the modern MLIPs \citep{batatia2022mace, neumann2024orb, qu2024importance, wood2026umafamilyuniversalmodels} to perform the structure relaxations, but that still leaves the magnetic moments themselves. Previous work \citep{Choudhary_2021, deng2023chgnet, xu2025spin} has tackled the this problem and instead use MLIPs trained on the converged magnetic moments to predict them. In particular, CHGNet \citep{deng2023chgnet} is trained on Materials Project and has demonstrated itself to be useful \citep{xu2025spin}. To use it, we simply load the pretrained \texttt{0.3.0} version the authors provide in their repository and evaluate it on our MP and GNoME test sets. On MP, CHGNet has an on-site MAE of $0.055 \;\mu_{B}$ and $0.060 \;\mu_{B}$ alongside a total mag. mom. MAE of $0.716 \;\mu_{B}$ and $0.472 \;\mu_{B}$, good accuracies both in and out of domain. A limitation of CHGNet is the choice to predict absolute values, thereby making it impossible for the model to distinguish between ferromagnetic and antiferromagnetic spin, but since the latter is only a small part of MP and GNoME, it remains the best model choice.

\subsection{Baselines}\label{app:baseline_scf:baseline}

Initialization of the smooth
valence density $\tilde{\rho}^{+}$ in a PAW DFT calculation is not meaningful
without the corresponding augmentation occupancies $\mathbf d_a^{+}$ that
determine the on-site density correction. This point was also raised by \citet{elsborg2026electrafi}. Furthermore, this does not include
the smooth spin-difference density $\tilde{\rho}^{-}$ or its corresponding
spin-difference augmentation occupancies $\mathbf d_a^{-}$ required by a
spin-polarized DFT calculation (\texttt{ISPIN=2}), which is used throughout
the Materials Project and is also standard in similar datasets. To evaluate the effects of including different components for initialization, we recalculated the test set of Materials Project used by \citep{elsborg2026electrafi} with different schemes. The results are shown in Table \ref{table:scf_exps}, and has several noteworthy aspects. 

As reflected by the additional 7 \% reduction observed for the magnetic subset relative to the non-magnetic subset of the evenly split test set, magnetic calculations can benefit even more from accurate initialization, owing to their generally slower convergence. Below the Oracle results, initialization with only the smooth valence density
$\tilde{\rho}^{+}$ represents the practically achievable setting corresponding
to previous approaches\citep{jorgensen2022equivariant, koker2024higher, elsborg2025electra, elsborg2026electrafi} when used as presented. Since these frameworks do not predict the augmentation occupancies
$\mathbf d_a^{+}$ or the spin-dependent components
$\tilde{\rho}^{-}$ and $\mathbf d_a^{-}$ that could lead to faster convergence, the relative non-magnetic reduction of 0.6\% is effectively just a default VASP run. 

The next two rows show that initializing the complete $+$ density channel,
i.e., $\tilde{\rho}^{+}$ together with $\mathbf d_a^{+}$, gives a 10\%
reduction on average for non-magnetic materials without any spin information, but fails otherwise. The spin-dependent components
$\tilde{\rho}^{-}$ and $\mathbf d_a^{-}$ alone are also insufficient for SCF step reduction. 

Finally, the last four rows show the benefits of initializing magnetic moments
together with either default or converged $+$-channel components
$(\tilde{\rho}^{+},\mathbf d_a^{+})$. The former corresponds to a typical
DFT calculation in the MP workflow, showing that a relaxation- or ML-derived
magnetic moment guess is beneficial at least for non-magnetic structures.
The latter demonstrates that a good spin state guess combined with
converged-accuracy $+$-channel components provides the second-best
initialization in the table. Thus, even without explicitly initializing
$\tilde{\rho}^{-}$ and $\mathbf d_a^{-}$, we can achieve approximately
40\% and 20\% SCF step reductions for non-magnetic and magnetic structures,
respectively. Between MPRelaxSet and CHGNet the difference is relatively
small.


\begin{table}[t]
\small
\centering
\setlength{\tabcolsep}{3.5pt}
\begin{tabular}{lccccrrrr}
\toprule
& \multicolumn{4}{c}{Initialization components}
& \multicolumn{2}{c}{vs. Default [\%]}
& \multicolumn{2}{c}{vs. Oracle [\%]} \\
\cmidrule(lr){2-5}
\cmidrule(lr){6-7}
\cmidrule(lr){8-9}

& \multicolumn{2}{c}{Smooth density}
& \multicolumn{2}{c}{PAW augmentation}
& & & & \\[-0.4em]
\cmidrule(lr){2-3}
\cmidrule(lr){4-5}

Initialization
    & $\tilde{\rho}^{+}$
    & $\tilde{\rho}^{-}$
    & $\mathbf d^{+}$
    & $\mathbf d^{-}$
    & Non-mag.
    & Mag.
    & Non-mag.
    & Mag. \\
\midrule

\multicolumn{9}{l}{\textit{\textbf{Baselines}}} \\[0.1em]

Default
& $\times$ & $\times$ & $\times$ & $\times$
& -- & -- & -- & -- \\

Oracle
& \cmark & \cmark & \cmark & \cmark
& \pos{49.0} & \pos{55.4}
& -- & -- \\

\addlinespace[0.55em]

\multicolumn{9}{l}{\textit{\textbf{Electronic component ablations}}} \\[0.1em]

Smooth valence only
& \cmark & $\times$ & $\times$ & $\times$
& \pos{0.5} & \nega{29.4}
& \nega{95.2} & \nega{189.9} \\

Spin channel only
& $\times$ & \cmark & $\times$ & \cmark
& \nega{9.7} & \nega{70.7}
& \nega{115.2} & \nega{282.3} \\

Smooth grids only
& \cmark & \cmark & $\times$ & $\times$
& \pos{10.2} & \nega{4.2}
& \nega{76.2} & \nega{133.3} \\

Augmentation only
& $\times$ & $\times$ & \cmark & \cmark
& \nega{14.3} & \nega{64.9}
& \nega{124.2} & \nega{269.2} \\

\addlinespace[0.55em]

\multicolumn{9}{l}{\textit{\textbf{Magnetic moment initialization only}}} \\[0.1em]

Oracle MagMom
& $\times$ & OM & $\times$ & OM
& \pos{12.0} & \nega{3.1}
& \nega{72.6} & \nega{130.8} \\

MPRelaxSet
& $\times$ & \mpmag & $\times$ & \mpmag
& \pos{9.3} & \nega{8.5}
& \nega{77.9} & \nega{142.9} \\

CHGNet
& $\times$ & \chgmag & $\times$ & \chgmag
& \pos{8.8} & \nega{5.7}
& \nega{78.9} & \nega{136.8} \\

\addlinespace[0.55em]

\multicolumn{9}{l}{\textit{\textbf{Smooth valence + augmentation initialization}}} \\[0.1em]

No spin init.
& \cmark & $\times$ & \cmark & $\times$
& \pos{13.5} & \nega{11.3}
& \nega{69.7} & \nega{149.3} \\

+ Oracle MagMom
& \cmark & OM & \cmark & OM
& \pos{47.6} & \pos{24.6}
& \nega{2.8} & \nega{68.9} \\

+ MPRelaxSet
& \cmark & \mpmag & \cmark & \mpmag
& \pos{42.4} & \pos{17.2}
& \nega{13.1} & \nega{85.6} \\

+ CHGNet
& \cmark & \chgmag & \cmark & \chgmag
& \pos{38.7} & \pos{20.4}
& \nega{20.2} & \nega{78.3} \\

\bottomrule
\end{tabular}

\caption{%
Paired SCF-step savings (\%) on the Materials Project test set, separated into
non-magnetic and magnetic structures.
The initialization components are the smooth spin-summed valence density
$\tilde{\rho}^{+}$, smooth spin-difference density $\tilde{\rho}^{-}$,
spin-summed PAW augmentation occupancies $\mathbf d^{+}$, and spin-difference
PAW augmentation occupancies $\mathbf d^{-}$, where $\mathbf d^{\pm}$ denotes
the collection of atom-wise occupancies $\{\mathbf d_a^{\pm}\}_a$.
Default denotes standard SAD initialization, while Oracle uses all converged
electronic components from the corresponding completed calculation.
\cmark{} denotes a converged component and $\times$ denotes SAD/default
initialization.
OM, \mpmag{} and \chgmag{} denote spin initialization from converged atomic
magnetic moments, MPRelaxSet, and CHGNet-predicted magnetic moments,
respectively.
Positive values indicate faster calculations and negative values indicate
slower calculations relative to the corresponding baseline.
}
\label{table:scf_exps}
\end{table}

\subsection{Spin-restricted DFT}\label{app:baseline_cf:ispin1}
To evaluate whether the benefits of machine learning-based initialization persist in the absence of spin coupling, we evaluate the non-magnetic Materials Project test set using the same computational parameters as previous work \citep{koker2024higher, elsborg2025electra, elsborg2026electrafi}, with the sole exception of setting \texttt{ISPIN=1}. The resulting SCF reductions are reported in Table~\ref{table:ispin1_scf_exps}. The results show that \texttt{ISPIN=1} calculations converge even faster than the \texttt{ISPIN=2} variants on the non-magnetic test set. While this approach sidesteps any discussion of the spin difference initialization, it is not applicable without prior knowledge of a structures magnetic behavior. Should the structure be magnetic according to our definition, the average number SCF steps increases to 44.28, almost twice as many as an SAD \texttt{ISPIN=2} calculation with 27.96 steps on average, while also converging to the wrong spin state. Since this is both physically incorrect for about 50\% of the MP database and also an inappropriate DFT approach, we deem this not a worth while direction to pursue for DFT initialization.


\begin{table}[h]
  \small
  \centering
  \begin{tabular}{lccc}
    \toprule
    Variant (\texttt{ISPIN=1]}) & Default (SAD) & Oracle & ChargE3Net$^{1, *}$ \\
    \midrule
    DFT Steps $\downarrow$ & 15.33 $\pm$ 8.02 & 9.21 $\pm$ 7.82 & 12.34 $\pm$ 8.69 \\
    DFT Time $\downarrow$ & 163.30 $\pm$ 326.32 s & 114.89 s $\pm$ 276.57 & 141.49 $\pm$ 298.70 s \\
    DFT Steps Saved $\uparrow$ & -- & 39.90 \% & 19.51 \% \\
    DFT Time Saved $\uparrow$ & -- & 29.64 \% & 13.35 \% \\
    \bottomrule
    \\
  \end{tabular}
  \caption{$^1$ \citet{koker2024higher}. $^*$ ChargE3Net is initialized with converged augmentation occupancies. DFT convergence analysis with different initializations. The tests were performed on the non-magnetic part of the MP test set.}
  \label{table:ispin1_scf_exps}
\end{table}

\clearpage
\section{PAW Augmentation and the AugNet Architecture}
\label{app:mace-aug}

\subsection{PAW Augmentation Occupancies as Covariant Targets}
\label{app:paw-occupancies}
In the PAW formalism, the spin-summed and spin-difference valence density
channels are represented as smooth pseudo-densities plus one-center corrections. The atom-centered correction is determined by the PAW setup and by a set of augmentation occupancy coefficients. In a VASP-style \texttt{CHGCAR}, these coefficients appear as augmentation occupancy blocks. For each atom $a$, the coefficients can be indexed schematically as
\begin{equation}
d^{LM,q}_{a,ij}, \qquad q\in\{+,-\},
\end{equation}
where $q=+$ denotes the spin-summed channel and $q=-$ the spin-difference
channel, $i$ and $j$ identify PAW partial-wave channels, and $L,M$ describe
the angular momentum channel of the coupled augmentation component.

This is the same structural object considered by the covariant Jacobi-Legendre PAW occupancy model of \citet{focassio2024covariant} and, for general non-magnetic materials, by EdenGNN~\citep{li2025efficient}. Compared to CJM, our AugNet model has two practical advantages. First, AugNet naturally extends to chemically diverse datasets containing many elements and PAW schemas, since the same backbone and equivariant readout are shared across atoms and the required output coefficients are selected according to the element-specific PAW schema. Second, the model can exploit the expressive nonlinear environment representation learned by MACE backbone rather than relying on a small fixed polynomial expansion. The trade-off is that AugNet contains more parameters and is less interpretable than a linear covariant expansion. For the task of accelerating general and diverse DFT calculations, AugNet is therefore a more practical model. 

\subsection{Augmentation Schemas and Equivariance}
Augmentation occupancies are not all rotational invariants. For a fixed $L$, the $(2L+1)$ components with $M=-L,\ldots,L$ transform together as an irreducible spherical tensor. Thus, if a structure is rotated, the target vector for each atom must rotate according to the corresponding Wigner representation. We demonstrate this experimentally further in appendix \ref{app:wigner}. This motivates representing the target as a direct sum of irreducible representations,
\begin{equation}
\mathbf d_a^q
=
\{d^{LM,q}_{a,ij}\}_{ijLM}
\in
\mathcal V_{s_a}
=
\bigoplus_L n_L^{(s_a)}D^L,
\qquad q\in\{+,-\},
\end{equation}
where $D^L$ denotes the $(2L+1)$-dimensional irrep of $SO(3)$ and
$n_L^{(s_a)}$ is the number of independent copies of angular channel $L$
for PAW schema $s_a=s(Z_a)$ associated with atom $a$.

The PAW setup determines which channels exist. In our implementation, all elements are mapped to one of five schema sizes:
\begin{equation}
s(Z_a) \in \{15,33,78,138,390\}.
\end{equation}
where $s(Z_a)$ is the number of valid augmentation coefficients for element $Z_a$. The corresponding irreducible representation decompositions are
\begin{align}
\label{eq:schemas1}
15  &: \quad 4\times 0\mathrm{e} + 2\times 1\mathrm{o} + 1\times 2\mathrm{e}, \\
33  &: \quad 6\times 0\mathrm{e} + 4\times 1\mathrm{o} + 3\times 2\mathrm{e}, \\
78  &: \quad 7\times 0\mathrm{e} + 6\times 1\mathrm{o} + 6\times 2\mathrm{e}
        + 2\times 3\mathrm{o} + 1\times 4\mathrm{e}, \\
138 &: \quad 9\times 0\mathrm{e} + 8\times 1\mathrm{o} + 10\times 2\mathrm{e}
        + 4\times 3\mathrm{o} + 3\times 4\mathrm{e}, \\
390 &: \quad 12\times 0\mathrm{e} + 12\times 1\mathrm{o} + 17\times 2\mathrm{e}
        + 12\times 3\mathrm{o} + 10\times 4\mathrm{e}
        + 4\times 5\mathrm{o} + 3\times 6\mathrm{e}.
        \label{eq:schemas5}
\end{align}
All targets are padded to dimension $390$, and a binary mask indicates which coefficients are valid for each atom.

\subsection{MACE backbone}
\label{app:mace}

AugNet uses MACE as the equivariant message-passing backbone. MACE constructs per-atom features by expanding local atomic environments in a basis of radial functions and spherical harmonics, then iteratively mixes these features through equivariant tensor products. The resulting node features transform as a direct sum of irreducible representations,
\begin{equation}
\mathbf h_a
\in
\bigoplus_{\ell=0}^{\ell_{\max}}
n_{\ell} D^\ell.
\end{equation}

In our implementation, the hidden irreps are specified by a width $w$ and maximum angular order $\ell_{\max}$,

\begin{equation}
\mathbf h_a
\in
w\times 0e
\oplus
w\times 1o
\oplus
\cdots
\oplus
w\times \ell_{\max}^{p_\ell},
\end{equation}

where $p_\ell=e$ for even $\ell$ and $p_\ell=o$ for odd $\ell$. The MACE interaction stack produces a sequence of equivariant node representations. We concatenate the node features from the interaction blocks before passing them to the PAW readout, so that the effective readout representation scales with both the hidden width and the number of interaction layers,

\begin{equation}
\mathbf h_a^{\mathrm{readout}}
=
\mathrm{concat}
\left(
\mathbf h_a^{(1)},\ldots,\mathbf h_a^{(T)}
\right).
\end{equation}

Here $T$ is the number of MACE interaction blocks. Increasing the hidden width increases the number of channels per irrep, while increasing the number of interaction blocks increases both the receptive field depth and the dimensionality of the representation passed to the PAW head.

The main expressivity knobs of the backbone are therefore the hidden width, the number of interaction blocks, the MACE correlation order, and the maximum angular order. In practice, width and depth primarily control general capacity, correlation controls the many-body order of the local expansion, and $\ell_{\max}$ controls the angular resolution of the equivariant features.

\subsection{Schema-agnostic Equivariant Readout}
\label{app:mace-aug-readout}

The output dimensionality and irrep content depend on the element-specific PAW schema (\ref{eq:schemas1}-\ref{eq:schemas5}) but they can be represented with a shared basis. AugNet therefore uses a shared readout across all schemas, using backbone
features up to angular order $L_{\mathrm{backbone}}$ and constructing
higher-order $L$ irreps through Clebsch-Gordan coupling in a fixed basis.

For an atom $a$, the schema is determined by its atomic number,

\begin{equation}
s_a = s(Z_a),
\end{equation}

with a maximum quantum number $L_a$. For PAW blocks $(i,j,L)$ with $L\leq L_{\mathrm{backbone}}$ of the backbone model, we can use a fully equivariant linear map to produce the outputs directly:

\begin{equation}
    \widehat{\Delta d}^{LM,q}_{a,ij}
    =
    \left[
        \mathcal W^{q}_{ijL}
        \left(\mathbf h_a^{\mathrm{readout}}\right)
    \right]_M,
    \qquad
    L\leq L_{\mathrm{backbone}},
\end{equation}

For $L>L_{\mathrm{backbone}}$, we first use an equivariant linear map to project the hidden representation to the projector space by $R$ coefficients for every slot $i$ in the maximal partial-wave basis $\{l_i \}^{n_{\max}}_{i=1}$:

\begin{equation}
    c^{(k)}_{a,i,m_i}
    =
    \left[
        \mathcal W_{\mathrm P}
        \left(\mathbf h_a^{\mathrm{readout}}\right)
    \right]_{i,k,m_i}.
\end{equation}

The corresponding PAW blocks are then built using an a Clebsch-Gordan contraction with a fixed basis to ensure that weights are shared across basis sets:

\begin{equation}
    \widehat{\Delta d}^{LM,q}_{a,ij}
=
\sum_{k=1}^{K} w_{(ij,L),k}
\sum_{m_i,m_j}
C^{LM}_{\ell_i m_i,\ell_j m_j}
c^{(k)}_{a,i,m_i}c^{(k)}_{a,j,m_j}.
\end{equation}

Finally, the atom-wise augmentation occupancy correction is constructed by
gathering the components specified by the schema:

\begin{equation}
    \widehat{\Delta\mathbf d}_a^q
    =
    \mathbf G_{s(Z_a)}
    \left(
        \{\widehat{\Delta d}^{LM,q}_{a,ij}\}_{ijLM}
    \right).
\end{equation}

\subsection{Reference baselines and $\Delta$-learning} \label{app:gates}

For the spin-summed channel, we use VASP's superposition-of-atomic-densities
(SAD) occupancies as the reference. Because the free-atom SAD reference is spherically symmetric, it is nonzero only for the $L=0$ channels, so higher-order
components therefore use a zero baseline. For the spin-difference channel, the
free-atom reference is not applicable, so we use a zero reference, making
$\Delta$-learning equivalent to direct prediction. We write
both cases as
\begin{equation}
    \hat{\mathbf d}_a^{q}
    =
    \mathbf d_a^{q,\mathrm{ref}}
    +
    \widehat{\Delta\mathbf d}_a^{q},
    \qquad
    \mathbf d_a^{q,\mathrm{ref}}
    =
    \begin{cases}
        \mathbf d_a^{+,\mathrm{SAD}}, & q=+,\\
        \mathbf 0, & q=-.
    \end{cases}
\end{equation}

The free-atom SAD references are only extracted once per PAW dataset with each extraction taking a few minutes at most. Adding the SAD allows for easy transfer between different PAW datasets, allowing the model to focus on higher-order components determined by the chemistry.

\subsection{Coefficient conventions}
\label{app:basis-transform}

Raw augmentation occupancies are stored in the PAW ordering. This ordering is organized by partial-wave pairs and angular channels. In contrast, e3nn expects coefficients grouped by irreducible representation. We therefore distinguish between two operations.

First, coefficients are permuted from the PAW channel ordering into e3nn grouped irrep ordering. Second, the real spherical harmonic convention used in the PAW representation is transformed into the real spherical harmonic convention used by e3nn. For a selected channel $q\in\{+,-\}$, we construct an orthogonal matrix $Q_L$ for each angular momentum $L$ such that
\begin{equation}
\mathbf d_{a,L}^{q,\mathrm{e3nn}}
=
Q_L\,
\mathbf d_{a,L}^{q,\mathrm{PAW}},
\end{equation}
The inverse transformation is
\begin{equation}
\mathbf d_{a,L}^{q,\mathrm{PAW}}
=
Q_L^\top
\mathbf d_{a,L}^{q,\mathrm{e3nn}}.
\end{equation}
with the transpose equal to the inverse because $Q_L$ is orthogonal. In implementation, $Q_L$ is obtained by evaluating both real spherical harmonic conventions on a deterministic set of points on the sphere and solving the least-squares basis alignment problem, followed by orthogonal projection.

Training is performed in the e3nn basis, which is the natural basis for the equivariant model. For evaluation and file writing, predictions are transformed back to the PAW basis. This also makes the reported PAW component metrics comparable to previous PAW occupancy parity plots.

\subsection{Training objective}
\label{app:mace-aug-loss}
AugNet is trained separately for the spin-summed ($+$) and spin-difference
($-$) augmentation channels. For a selected channel $q\in\{+,-\}$, the
dataset provides target coefficients $\mathbf d_a^q$ and a schema mask $m$. The loss is a masked coefficient space loss. For the mean squared error case,
\begin{equation}
\mathcal L
=
\frac{
    \sum_{a,\alpha}
    m_{a,\alpha}
    \left(
        \hat d^q_{a,\alpha}
        -
        d^q_{a,\alpha}
    \right)^2
}{
    \sum_{a,\alpha}
    m_{a,\alpha}
},
\end{equation}
where $\alpha$ indexes the packed $(ij,L,M)$ coefficients. We note here that VASP data carries the \texttt{LMAXMIX} parameter, dictating the maximum $L$ that is both used by the density mixer but also written to the CHGCAR. This means that if $\texttt{LMAXMIX}=2$, all augmentation occupancies for a given atom above that will be set to 0, regardless of the schema it carries. That means that the masked coefficient loss will also extend to mask out any 0's set by \texttt{LMAXMIX}. 


\subsection{Evaluation metrics}
\label{app:mace-aug-metrics}

For a selected augmentation channel $q\in\{+,-\}$, errors are computed over
all valid PAW coefficients. In the e3nn basis, the masked coefficient MAE, RMSE, and MaxAE are
\begin{align}
\mathrm{MAE}
&=
\frac{1}{N_{\mathrm{coeff}}}
\sum_{a,\alpha}
m_{a,\alpha}
\left|
\hat d^q_{a,\alpha}-d^q_{a,\alpha}
\right|,
\\
\mathrm{RMSE}
&=
\left[
\frac{1}{N_{\mathrm{coeff}}}
\sum_{a,\alpha}
m_{a,\alpha}
\left(
\hat d^q_{a,\alpha}-d^q_{a,\alpha}
\right)^2
\right]^{1/2}.
\end{align}

\subsection{AugNet Accuracy and Transfer}
\label{app:augnet-accuracy-transfer}

Table~\ref{table:augnet_results} reports spin-summed augmentation occupancy
$\mathbf d^{+}$ prediction errors as the
Materials Project training set is increased from 1k structures to the full
training set. The full model reaches a mean per-structure MAE/RMSE of
$0.0041/0.0118$ on the MP test set and $0.0062/0.0262$ on the OOD GNoME
test set. Accuracy improves consistently with training set size on MP, while
the OOD results begin to saturate at larger dataset sizes. The unusually large
GNoME MaxAE originates from a small number of extreme coefficient outliers:
after excluding the ten largest errors, MaxAE falls from $366.60$ to $0.867$.
\begin{table*}[ht!]
\centering
\caption{
Physical augmentation occupancy prediction errors versus training-set size
on the full MP and GNoME test sets. RMSE and MAE are means over
per-structure values, while MaxAE is the largest single-coefficient error.
MaxAE$_{\mathrm{top\,11}}$ reports the largest error after excluding the ten
most extreme coefficients.
}
\label{table:augnet_results}

\small
\setlength{\tabcolsep}{5pt}
\renewcommand{\arraystretch}{0.95}

\begin{tabular*}{\textwidth}{@{\extracolsep{\fill}}clrrrr}
\toprule
Dataset
& Training set
& RMSE $\downarrow$
& MAE $\downarrow$
& MaxAE $\downarrow$
& MaxAE$_{\mathrm{top\,11}}\downarrow$
\\
\midrule

\multirow{4}{*}{\textbf{MP}}
& 1k
& 0.0340 $\pm$ 0.0206
& 0.0120 $\pm$ 0.0065
& 4.504
& 1.487
\\
& 10k
& 0.0183 $\pm$ 0.0113
& 0.0064 $\pm$ 0.0034
& 2.829
& 0.796
\\
& 50k
& 0.0130 $\pm$ 0.0096
& 0.0046 $\pm$ 0.0028
& 1.931
& 0.664
\\
& Full
& \textbf{0.0118 $\pm$ 0.0091}
& \textbf{0.0041 $\pm$ 0.0026}
& 1.114
& 0.646
\\

\midrule

\multirow{4}{*}{\textbf{GNoME}}
& 1k
& 0.0464 $\pm$ 0.3916
& 0.0120 $\pm$ 0.0287
& 366.85
& 1.724
\\
& 10k
& 0.0302 $\pm$ 0.3915
& 0.0078 $\pm$ 0.0284
& 366.58
& 0.868
\\
& 50k
& 0.0271 $\pm$ 0.3915
& 0.0065 $\pm$ 0.0284
& 366.58
& 0.868
\\
& Full
& \textbf{0.0262 $\pm$ 0.3916}
& \textbf{0.0062 $\pm$ 0.0284}
& 366.60
& 0.867
\\

\bottomrule
\end{tabular*}
\end{table*}
Two prior models directly target the same PAW augmentation occupancy
object. EdenGNN~\citep{li2025efficient} predicts them for general non-magnetic materials and reports an MAE of $0.0085$ on MP structures recomputed with its own VASP settings, but does not use them for SCF initialization. CJM~\citep{focassio2024covariant} is trained and evaluated on a
much narrower dataset containing ab initio molecular dynamics configurations of
$\mathrm{MoS}_2$ in the 1H and 1T phases and intermediate geometries, and
reports an MAE/RMSE of $0.0130/0.0459$. This provides a useful external
reference for the scale of coefficient space errors, although it is not a
strict matched benchmark because the datasets and aggregation procedures differ.
Compared with this system-specific reference, AugNet reaches errors on the same
or lower scale while operating across chemically diverse structures, elements,
and PAW schemas.

We further test transfer directly on the $\mathrm{MoS}_2$ dataset of
\citet{focassio2024covariant} (Table~\ref{table:sanvito-comp}). This setting
changes the underlying Mo PAW dataset relative to Materials Project and
therefore changes the target augmentation representation itself. Consequently,
the MP-pretrained model does not transfer zero-shot. However, AugNet adapts
readily to the new PAW setup: training from scratch reaches an RMSE of $0.0258$,
already below the $0.0459$ reported for CJM, while full fine-tuning of the
pretrained model reaches $0.0115$. Fine-tuning only the PAW readout requires
only $73$k trainable parameters and reaches an RMSE of $0.0400$. These results
indicate that AugNet generalizes well within a fixed PAW representation and can
be efficiently adapted when the underlying PAW setup changes.

\begin{table}[t!]
\centering
\small
\setlength{\tabcolsep}{5pt}

\begin{tabular}{lrrrrr}
\toprule
Model
& Steps
& Trainable params.
& MAE $\downarrow$
& RMSE $\downarrow$
& MaxAE $\downarrow$
\\
\midrule

CJM~\citep{focassio2024covariant}
& 1k
& 1,758
& 0.0130
& 0.0459
& 0.9137
\\

\midrule

AugNet, zero-shot
& --
& --
& 0.3854
& 1.3439
& 14.9708
\\

AugNet, from scratch
& 1k
& 3M
& \textbf{0.0010}
& 0.0258
& 0.4969
\\

\addlinespace[0.3em]

\multirow{2}{*}{AugNet, head fine-tune}
& 1k
& 73k
& 0.1801
& 0.7039
& 8.2880
\\
&
10k
& 73k
& 0.0122
& 0.0400
& 0.9490
\\

\addlinespace[0.3em]

\multirow{2}{*}{AugNet, full fine-tune}
& 1k
& 3M
& 0.0157
& 0.0352
& 0.8535
\\
&
10k
& 3M
& 0.0051
& \textbf{0.0115}
& \textbf{0.2353}
\\

\bottomrule
\end{tabular}

\caption{
Transfer of MP-pretrained AugNet to the $\mathrm{MoS}_2$ PAW setup of
\citet{focassio2024covariant}. Their calculations use a different Mo PAW
dataset from Materials Project, so the target augmentation representation
changes and direct zero-shot transfer is not expected. ``Trainable params.''
denotes the number of parameters optimized during adaptation.
}
\label{table:sanvito-comp}
\end{table}

\clearpage
\section{Magnetic Initialization Development}
\label{app:magnetic-init}

To construct a fully learned spin initialization, we extend both ELECTRAFI
and AugNet to the spin-difference components of the PAW density:
spin-ELECTRAFI predicts the smooth spin-difference density
$\tilde{\rho}^{-}$ on the plane-wave grid, and spin-AugNet predicts the
corresponding spin-difference augmentation occupancies $\mathbf d_a^{-}$.
Together with the spin-summed components
$(\tilde{\rho}^{+},\mathbf d_a^{+})$, these predictions provide all
structure-dependent density components required for direct spin-polarized
initialization. Architectural details of AugNet are given in
Appendix~\ref{app:mace-aug} and those of ELECTRAFI in
\citet{elsborg2026electrafi}.

\paragraph{Spin-ELECTRAFI.}\label{app:spin-electrafi}
We adapt the ELECTRAFI model of \citet{elsborg2026electrafi} to additionally
predict the smooth spin-difference density
$\tilde{\rho}^{-}
=
\tilde{\rho}_{\uparrow}
-
\tilde{\rho}_{\downarrow}$. In the process, several computational inefficiencies of the original implementation were removed, reducing training and inference time without altering the numerics of the model.

The simplest extension within the ELECTRAFI ansatz is to allocate a second
set of signed weights $w^{-}$ to the spin-difference density while sharing
the Gaussian centers and covariances with the spin-summed smooth valence density
$\tilde{\rho}^{+}$ The
spin-difference weights are predicted analogously to the spin-summed weights.
For Gaussian $\mathcal N^{(j)}$,
\begin{equation}
  w^{(j),-}
  =
  \tanh\!\big(s^{(j),-}\big),
  \qquad
  s^{(j),-}
  =
  f_{w,-}\!\big(S^{(j)}\big),
\end{equation}
where $S \in \mathbb{R}^{N \times C}$ are the scalar outputs of the ELECTRAFI backbone for $N$ atoms and channel width $C$, and $f_{w,-}$ is a multilayer perceptron (MLP) with the same architecture
as the spin-summed weight MLP $f_{w,+}$. The Gaussian centers $\boldsymbol{\mu}^{(j)}$ and covariances $\boldsymbol{\Sigma}^{(j)}$ are predicted as in the original model. Both densities are then assembled through the analytic Fourier transform of the Gaussian ansatz followed by an inverse FFT~\citep{elsborg2026electrafi}:
\begin{equation}
  \hat{\tilde{\rho}}^{+}(\mathbf G)
  =
  \sum_{j=1}^{N_\mathcal{N}}
  w^{(j),+}
  \exp\!\Big[
      -\tfrac12\mathbf G^\top
      \boldsymbol{\Sigma}^{(j)}
      \mathbf G
  \Big]
  e^{-i\mathbf G\cdot\boldsymbol{\mu}^{(j)}},
  \qquad
  \hat{\tilde{\rho}}^{+}(\mathbf r)
  =
  \mathrm{IFFT}
  \big[
      \hat{\tilde{\rho}}^{+}(\mathbf G)
  \big](\mathbf r),
\end{equation}

\begin{equation}
  \hat{\tilde{\rho}}^{-}(\mathbf G)
  =
  \sum_{j=1}^{N_\mathcal{N}}
  w^{(j),-}
  \exp\!\Big[
      -\tfrac12\mathbf G^\top
      \boldsymbol{\Sigma}^{(j)}
      \mathbf G
  \Big]
  e^{-i\mathbf G\cdot\boldsymbol{\mu}^{(j)}},
  \qquad
  \hat{\tilde{\rho}}^{-}(\mathbf r)
  =
  \mathrm{IFFT}
  \big[
      \hat{\tilde{\rho}}^{-}(\mathbf G)
  \big](\mathbf r).
\end{equation}

Sharing the Gaussian centers and covariances between the two channels is physically motivated. In collinear spin-polarized DFT the spin-resolved densities are non-negative, so the spin-difference density is bounded pointwise by the total density, $|\tilde{\rho}^{-}(\mathbf r)|
\le
\tilde{\rho}^{+}(\mathbf r)$: magnetization can only exist where charge exists. Moreover, the net spin polarization is carried by the same partially filled, localized orbitals that dominate the total density around magnetic atoms, so the spatial support and characteristic length scales of
$\tilde{\rho}^{-}$ are inherited from $\tilde{\rho}^{+}$, and the two fields differ primarily in sign and magnitude. A Gaussian that is prominent in the spin-summed density is therefore also the natural carrier of any spin difference in the same region, whereas a Gaussian with negligible spin-summed weight should carry no magnetization. The shared basis encodes this structure directly: the geometry of the expansion is fixed by the spin-summed density and only signed magnitudes are learned per channel. This acts as a physical regularizer on $\tilde{\rho}^{-}$, avoids predicting a second set of centers and covariances, and for non-magnetic structures reduces to training $w^{-}\to0$.

Because the spin-difference head reuses the backbone and Gaussian parameters, the backbone continues to receive the clean geometric signal of the spin-summed density while learning the comparatively sparse magnetization density, which stabilizes joint training. The cost is nearly two readout passes and a correspondingly more expensive backward pass per optimization step. The loss is the sum of a spin-summed density term, given by the normalized MAE (NMAE) of \citet{jorgensen2022equivariant},
\begin{equation}
\label{app:eq:elec_loss}
\mathcal L_{\tilde{\rho}^{+}}
=
\operatorname{NMAE}
\big(
    \hat{\tilde{\rho}}^{+},
    \tilde{\rho}_{\mathrm{ref}}^{+}
\big)
=
\frac{
    \int_{\Omega}
    \left|
        \tilde{\rho}_{\mathrm{ref}}^{+}(\mathbf r)
        -
        \hat{\tilde{\rho}}^{+}(\mathbf r)
    \right|\,dV
}{
    \int_{\Omega}
    \tilde{\rho}_{\mathrm{ref}}^{+}(\mathbf r)\,dV
}.
\end{equation}
and a spin-difference term that distinguishes magnetic from non-magnetic structures,
\begin{equation}
\mathcal L_{\tilde{\rho}^{-}}
=
\begin{cases}
\dfrac{
    \int_{\Omega}
    \left|
        \tilde{\rho}_{\mathrm{ref}}^{-}(\mathbf r)
        -
        \hat{\tilde{\rho}}^{-}(\mathbf r)
    \right|\,dV
}{
    \int_{\Omega}
    \left|
        \tilde{\rho}_{\mathrm{ref}}^{-}(\mathbf r)
    \right|\,dV
},
& \text{magnetic},
\\[3ex]
\displaystyle
\int_{\Omega}
\left|
    \tilde{\rho}_{\mathrm{ref}}^{-}(\mathbf r)
    -
    \hat{\tilde{\rho}}^{-}(\mathbf r)
\right|\,dV,
& \text{non-magnetic}.
\end{cases}
\end{equation}
Following~\citet{koker2024higher,elsborg2026electrafi}, we classify a structure
as magnetic when
$M_{\mathrm{abs}}
=
\int_\Omega|\tilde{\rho}_{\mathrm{ref}}^{-}|\,dV
>
M_{\min}=0.1$. The total loss is
\begin{equation}
\mathcal L
=
\mathcal L_{\tilde{\rho}^{+}}
+
\lambda_{\mathrm{spin}}
\mathcal L_{\tilde{\rho}^{-}},
\qquad
\lambda_{\mathrm{spin}}=0.2.
\end{equation}
The case distinction is necessary because, for non-magnetic structures, the NMAE denominator approaches zero and the loss term diverges. We also trained with a plain MAE loss for both channels, but this did not
yield a balanced contribution from $\tilde{\rho}^{+}$ and
$\tilde{\rho}^{-}$ and degraded performance.

Analogously to the spin-summed density head, we normalize the spin-difference
readout by the net magnetic moment to ensure well-behaved grid predictions. Unlike the number of valence electrons, however, this quantity is not available prior to the DFT calculation. We therefore normalize with the ground-truth moment $M_{\mathrm{DFT}}$ during training and substitute the CHGNet-predicted moment $\hat M_{\mathrm{CHGNet}}$ as a surrogate at inference. This approach works well with the exception of antiferromagnetic materials, a notoriously difficult spin state for DFT whose vanishing net moment cannot be resolved by CHGNet. We also attempted to omit the normalization entirely, but this rendered the training dynamics too unstable for long training runs.

Joint training increases the cost by roughly $2.5\times$: whereas spin-summed density training on MP for five epochs takes $2.5$ days on a single NVIDIA H200 GPU, joint training takes approximately one week. 

\paragraph{Spin-AugNet.}\label{app:augnet:spin}
Spin-AugNet uses the same equivariant architecture as AugNet with
spin-difference augmentation occupancies $\mathbf d_a^{-}$ as targets.
As described in Appendix~\ref{app:gates}, the free-atom SAD reference is not
applicable to the spin-difference channel, so we use
$\mathbf d_a^{-,\mathrm{ref}}=\mathbf 0$. Consequently,
\begin{equation}
    \hat{\mathbf d}_a^{-}
    =
    \widehat{\Delta\mathbf d}_a^{-},
\end{equation}
making the shared $\Delta$-learning formulation equivalent to direct
prediction for spin-AugNet.

\clearpage
\section{Detailed End-to-End DFT Results}
\label{app:end-to-end-results}

Table~\ref{table:comparison_SCF} (Tables \ref{table:comparison_SCF_mag} and \ref{table:comparison_SCF_nonmag} for the magnetic and non-magnetic subsets) reports the complete numerical results underlying the end-to-end comparison in
Figure~\ref{fig:finalfig}. In addition to total wall time, we report
density prediction accuracy, SCF iterations, DFT execution time, and ML
initialization overhead. The Default calculation uses the standard VASP
initialization, while Oracle uses the corresponding converged electronic components
$(\tilde{\rho}^{+},\mathbf d^{+},\tilde{\rho}^{-},\mathbf d^{-})$ as
initialization and therefore represents an empirical upper bound
on the achievable acceleration under the same DFT settings.

\begin{table*}[h!]
\centering
\caption{%
Detailed comparison of reference-free ML PAW initializations and resulting DFT
performance on the test sets of MP and GNoME. Total time includes both
ML initialization and DFT execution. For both CNEI-EFI and CNEI-C3Net, we add
the time it takes to evaluate spin-ELECTRAFI (0.24s/0.15s) and spin-AugNet (0.05s both)
as well as CHGNet (0.03s both) to ELECTRAFI and ChargE3Net.
}
\label{table:comparison_SCF}
\footnotesize
\setlength{\tabcolsep}{7pt}
\renewcommand{\arraystretch}{1.05}
\begin{tabular*}{\textwidth}{@{\extracolsep{\fill}}llcccc}
\toprule
Dataset & Metric & Default & Oracle & CNEI-EFI & CNEI-C3Net \\
\midrule
\multirow{8}{*}{\textbf{MP}}
& $\tilde{\rho}^{+}$ NMAE $\downarrow$            & --        & --        & 0.58\%             & 0.54\%              \\
& ML init time $\downarrow$    & --        & --        & $(0.24+0.37)$ s    & $(78.73+0.37)$ s    \\
& SCF steps $\downarrow$       & 22.05     & 10.41     & 19.05              & 18.36               \\
& DFT time $\downarrow$        & 623.84 s  & 302.04 s  & 529.43 s           & 506.16 s            \\
& Total time $\downarrow$      & 623.84 s  & 302.04 s  & 530.04 s           & 585.26 s            \\
\addlinespace[0.2em]
& SCF steps saved $\uparrow$   & --        & 52.78\%   & 13.62\%             & 16.76\%             \\
& DFT time saved $\uparrow$    & --        & 51.58\%   & 15.13\%            & 18.86\%             \\
& \textbf{Total time saved $\uparrow$}
                               & --        & \textbf{51.58\%} & \textbf{15.04\%}   & \textbf{6.18\%}  \\
\midrule
\multirow{8}{*}{\textbf{GNoME}}
& $\tilde{\rho}^{+}$ NMAE $\downarrow$            & --        & --        & 0.93\%             & 0.69\%              \\
& ML init time $\downarrow$    & --        & --        & $(0.15+0.28)$ s    & $(33.28+0.28)$ s    \\
& SCF steps $\downarrow$       & 16.30     & 7.89     & 11.87              & 11.45               \\
& DFT time $\downarrow$        & 188.99 s  & 112.59 s  & 141.00 s           & 140.46 s            \\
& Total time $\downarrow$      & 188.99 s  & 112.59 s  & 141.43 s           & 174.02 s            \\
\addlinespace[0.2em]
& SCF steps saved $\uparrow$   & --        & 51.63\%   & 27.17\%             & 29.79\%             \\
& DFT time saved $\uparrow$    & --        & 40.43\%   & 25.39\%            & 25.68\%             \\
& \textbf{Total time saved $\uparrow$}
                               & --        & \textbf{40.43\%} & \textbf{25.17\%}   & \textbf{7.92\%}  \\
\bottomrule
\end{tabular*}
\end{table*}

\begin{table*}[h!]
\centering
\caption{%
The magnetic subset counterpart of table~\ref{table:comparison_SCF}. The spin-ELECTRAFI model measures (0.24s/0.15s).
}
\label{table:comparison_SCF_mag}
\footnotesize
\setlength{\tabcolsep}{7pt}
\renewcommand{\arraystretch}{1.05}
\begin{tabular*}{\textwidth}{@{\extracolsep{\fill}}llcccc}
\toprule
Dataset & Metric & Default & Oracle & CNEI-EFI& CNEI-C3Net \\
\midrule
\multirow{8}{*}{\textbf{MP}}
& $\tilde{\rho}^{+}$ NMAE $\downarrow$                    & --       & --          & 0.67 \%             & 0.78 \%   \\
& ML Init Time $\downarrow$          & --       & --          & (0.24 + 0.37) s     & (87.25 + 0.37) s   \\
& SCF steps $\downarrow$       & 27.82     & 12.43     & 24.97              & 24.48               \\
& DFT time $\downarrow$        & 882.45 s  & 377.16 s  & 777.63 s           & 744.13 s            \\
& Total time $\downarrow$      & 882.45 s  & 377.16 s  & 778.24 s           & 831.75 s            \\
\addlinespace[0.2em]
& SCF steps saved $\uparrow$   & --        & 55.30\%   & 10.22\%             & 11.98\%             \\
& DFT time saved $\uparrow$    & --        & 57.26\%   & 11.88\%            & 15.67\%             \\
& \textbf{Total time saved $\uparrow$}
                               & --        & \textbf{57.26\%} & \textbf{11.81\%}   & \textbf{5.75\%}  \\
\midrule
\multirow{8}{*}{\textbf{GNoME}}
& $\tilde{\rho}^{+}$ NMAE $\downarrow$                    & --       & --   & 1.01 \%        & 0.92 \%   \\
& ML Init Time $\downarrow$          & --       & --   & (0.18+0.31) s  & (44.65 + 0.31) s   \\
& SCF steps $\downarrow$       & 22.60     & 10.06     & 14.69              & 14.49               \\
& DFT time $\downarrow$        & 344.88 s  & 190.82 s  & 238.26 s           & 238.22 s            \\
& Total time $\downarrow$      & 344.88 s  & 190.82 s  & 238.75 s           & 283.18 s            \\
\addlinespace[0.2em]
& SCF steps saved $\uparrow$   & --        & 55.47\%   & 34.99\%             & 35.88\%             \\
& DFT time saved $\uparrow$    & --        & 44.67\%   & 30.91\%            & 30.93\%             \\
& \textbf{Total time saved $\uparrow$}
                               & --        & \textbf{44.67\%} & \textbf{30.77\%}   & \textbf{17.89\%}  \\
\bottomrule
\end{tabular*}
\end{table*}

\begin{table*}[t!]
\centering
\caption{%
The non-magnetic subset counterpart of table~\ref{table:comparison_SCF}. The spin-ELECTRAFI model measures (0.17s/0.11s) on these subsets.
}
\label{table:comparison_SCF_nonmag}
\footnotesize
\setlength{\tabcolsep}{7pt}
\renewcommand{\arraystretch}{1.05}
\begin{tabular*}{\textwidth}{@{\extracolsep{\fill}}llcccc}
\toprule
Dataset & Metric & Default & Oracle & CNEI-EFI & CNEI-C3Net \\
\midrule
\multirow{8}{*}{\textbf{MP}}
& $\tilde{\rho}^{+}$ NMAE $\downarrow$                    & --       & --    & 0.55 \%             & 0.50 \%   \\
& ML Init Time $\downarrow$          & --       & --      & (0.17+0.30) s     & (72.11+0.30) s   \\
& SCF steps $\downarrow$       & 16.83     & 8.58     & 13.68              & 12.80               \\
& DFT time $\downarrow$        & 389.41 s  & 233.95 s  & 304.45 s           & 290.45 s            \\
& Total time $\downarrow$      & 389.41 s  & 233.95 s  & 304.92 s           & 362.86 s            \\
\addlinespace[0.2em]
& SCF steps saved $\uparrow$   & --        & 49.01\%   & 18.71\%             & 23.91\%             \\
& DFT time saved $\uparrow$    & --        & 39.92\%   & 21.82\%            & 25.41\%             \\
& \textbf{Total time saved $\uparrow$}
                               & --        & \textbf{39.92\%} & \textbf{21.70\%}   & \textbf{6.82\%}  \\
\midrule
\multirow{8}{*}{\textbf{GNoME}}
& $\tilde{\rho}^{+}$ NMAE $\downarrow$                    & --      & --         & 0.88 \%        & 0.59 \%   \\
& ML Init Time $\downarrow$          & --      & --         & (0.11+0.24) s  & (28.29+0.24) s   \\
& SCF steps $\downarrow$       & 13.48     & 6.91     & 10.61              & 10.08               \\
& DFT time $\downarrow$        & 119.11 s  & 77.52 s  & 97.40 s           & 96.64 s            \\
& Total time $\downarrow$      & 119.11 s  & 77.52 s  & 97.75 s           & 125.17 s            \\
\addlinespace[0.2em]
& SCF steps saved $\uparrow$   & --        & 48.75\%   & 21.29\%             & 25.22\%             \\
& DFT time saved $\uparrow$    & --        & 34.92\%   & 18.22\%            & 18.86\%             \\
& \textbf{Total time saved $\uparrow$}
                               & --        & \textbf{34.92\%} & \textbf{17.93\%}   & \textbf{-5.09\%}  \\
\bottomrule
\end{tabular*}
\end{table*}
\clearpage
\section{Equivariance of PAW augmentation occupancies}
\label{app:wigner}

For each atom $a$ and channel $q\in\{+,-\}$, VASP stores the PAW
augmentation occupancies as blocks
$\mathbf d_a^{(ij,L),q}\in\mathbb{R}^{2L+1}$ with components
$d^{LM,q}_{a,ij}$, $M=-L,\ldots,L$, packed into the augmentation occupancy
vector $\mathbf d_a^q$. The allowed channels are determined entirely by the
partial-wave angular momenta $(\ell_i,\ell_j)$ and truncated by
\texttt{LMAXMIX}. The allowed channels are determined
entirely by the partial-wave angular momenta $(l_i,l_j)$ and truncated by
\texttt{LMAXMIX}.

\paragraph{Rotation law.}
Under a rigid rotation $R\in SO(3)$ of the crystal, each block transforms as

\begin{equation}
\mathbf d_a^{(ij,L),q}(R)
=
Q_L^\top
D^L_{\mathrm{e3nn}}(R)
Q_L
\mathbf d_a^{(ij,L),q},
\qquad q\in\{+,-\},
\label{eq:rotationlaw}
\end{equation}

where $D^L_{\mathrm{e3nn}}$ is the real Wigner $D$-matrix and $Q_L$ is a fixed
change of basis between the VASP and e3nn spherical harmonic conventions.
Consequently, the augmentation occupancies form a direct sum of irreducible
$SO(3)$ representations and provide natural equivariant prediction targets.

\paragraph{Verification.}
We verified Equation~\ref{eq:rotationlaw} using 242 rigidly rotated VASP
calculations of \texttt{mp-1069193}. For each rotation, the occupancies predicted
from the identity calculation using Equation~\ref{eq:rotationlaw} were compared to
those written by VASP. The relative error

\begin{equation}
\varepsilon_a
=
\left(
\frac{
\sum_R
\left\|
\hat{\mathbf d}_a^q(R)-\mathbf d_a^q(R)
\right\|^2
}{
\sum_R
\left\|
\mathbf d_a^q(R)
\right\|^2
}
\right)^{1/2}.
\end{equation}

was below $10^{-6}$ for every atomic site (Table~\ref{tab:wigner_results}),
matching the numerical precision of the printed \texttt{CHGCAR} values. Using
the transpose representation or an incorrect partial-wave ordering increased the
error by approximately six orders of magnitude.

\begin{table}[h]
\centering
\begin{tabular}{lcc}
\toprule
Site & Partial waves & $\varepsilon$ \\
\midrule
1 & $(2,2,0,0,1,1)$ & $4.6\times10^{-7}$\\
2--5 & $(0,0,1,1)$ & $1.1$--$10.1\times10^{-7}$\\
\bottomrule
\end{tabular}
\caption{Parameter-free verification of the equivariant transformation law over
241 held-out rotations.}
\label{tab:wigner_results}
\end{table}

The PAW augmentation occupancies therefore provide an exact equivariant
coefficient representation of the on-site PAW augmentation correction, and the results show that they can be converted losslessly between
the VASP and e3nn conventions via the fixed matrices $Q_L$, allowing
$E(3)$-equivariant neural networks to predict augmentation occupancies in their
natural irreducible basis.

\clearpage
\section{Experiment Setup and Hyperparameters}\label{app:hyperparameters}

\subsection{Experimental Hardware}\label{app:hyperparamters:hardware}
All VASP experiments were conducted using the same Intel Xeon E5-2650 2.20GHz Broadwell CPUs and parallelized across 24 CPU cores using 256 GB of RAM. Machine learning models were trained on a mixture of NVIDIA A100 and H200 GPUs in single-GPU training. However, all inference timings were measured using A100 GPUs.

\subsection{AugNet}\label{app:hyperparameters:augnet}
\begin{table}[h]
\small
\centering
\begin{tabular}{lll}
\toprule
\textbf{Group} & \textbf{Hyperparameter} & \textbf{Value} \\
\midrule
\multirow{7}{*}{Backbone}
 & Hidden width & 64 \\
 & Max. spherical order $\ell_{\max}$ & 3 \\
 & Cutoff radius $r_{\max}$ (\AA) & 6.0 \\
 & Interaction layers & 2 \\
 & Correlation order & 3 \\
 & Avg. number of neighbors & 64.3 \\
 & Tensor product for high $L$ & Yes \\
 
\midrule

\multirow{3}{*}{Readout head}
 & Projector rank & 64 \\
 & Block mixing & Yes \\
 & Linear readout up to $L$ & 3 (CG coupling for $L \geq 4$) \\
 
\midrule

\multirow{2}{*}{Target}
 & Spin-summed ($+$) & $\Delta$ from SAD reference \\
 & Spin-difference ($-$) & Zero reference (direct prediction) \\
\midrule
\multirow{9}{*}{Optimization}
 & Optimizer & AdamW \\
 & Learning rate & $1\times10^{-2}$ \\
 & Backbone LR multiplier & 0.1 \\
 & Weight decay & $1\times10^{-6}$ \\
 & Gradient clipping & None \\
 & Epochs & 5 \\
 & Batch size & 1 \\
 & Precision & FP32 \\
 & Loss & MSE \\
\bottomrule
\end{tabular}
\caption{Hyperparameters used for training AugNet and spin-AugNet. The
spin-summed and spin-difference models share all architectural and optimization
settings and differ only in the predicted channel and reference baseline.}
\label{tab:hyperparams}
\end{table}

\newpage
\subsection{ELECTRAFI}\label{app:hyperparameters:electrafi}
\begin{table}[ht!]
\small
\centering
\begin{tabular}{lll}
\toprule
Group & Parameter & Value \\
\midrule
\multirow{13}{*}{Backbone (EScAIP)} &
Layers                         & $2$                      \\
& Hidden size                  & $256$                    \\
& Attention heads              & $32$                     \\
& Atom embedding size          & $128$                    \\
& Edge distance embedding      & $512$ (expansion $600$)  \\
& Node direction embedding     & $256$ (expansion $13$)   \\
& FFN hidden multiplier        & $2$                      \\
& Activation                   & GELU                     \\
& Normalization                & LayerNorm                \\
& Dropout / stochastic depth   & $0$                      \\
& Max neighbours               & $300$                    \\
& Batch size                   & $1$                       \\
& Master units                 & $2160$                    \\
\midrule
\multirow{7}{*}{Density representation}
& Gaussians per electron       & $120$                    \\
& Signed weights               & \texttt{tanh\_softplus}  \\
& Weight magnitude cap         & $50$                     \\
& Gaussian width scale range   & $[0.01,\ 25]$            \\
& Gaussian width floor         & $1\times10^{-4}$         \\
& Renormalization floor        & $10^{-3}\sum|w|$         \\
& Plane-wave grid              & $128^3$                  \\
\midrule
\multirow{3}{*}{Spin-difference channel}
& Spin loss weight                   & $0.2$                    \\
& Magnetic threshold $M_{\mathrm{abs}}$ & $0.1\,\mu_B$     \\
& Spin loss warm-up                  & $500$ steps              \\
\midrule
\multirow{14}{*}{Optimization} &
Epochs                       & 5                        \\
& Optimizer                    & Muon $+$ AdamW (AMSGrad)          \\
& Learning rate (AdamW)        & $3\times10^{-4}$         \\
& Learning rate (Muon)         & $3\times10^{-3}$         \\
& Final learning rate          & $1\times10^{-4}$         \\
& LR decay                     & $\gamma=0.7$ per epoch   \\
& Weight decay                 & $0$                      \\
& Muon momentum (Nesterov)     & $0.95$                   \\
& Newton--Schulz steps         & $5$                      \\
& Gradient clipping            & $1.0$ (norm)             \\
& Loss                         & Equation~\ref{app:eq:elec_loss}                     \\
& Loss-spike skip threshold    & $50\times$ EMA           \\
& Training Time Rotations      & True                     \\
& Precision                    & FP32                     \\
\bottomrule \\
\end{tabular}
\caption{Hyperparameters for the ELECTRAFI and spin-ELECTRAFI models,
mirroring the choices in~\citet{elsborg2026electrafi}.}
\end{table}

\clearpage



\end{document}